\documentclass{aa}
\pdfoutput=1
\usepackage{amsmath,bm} 
\usepackage[varg]{txfonts}
\usepackage{natbib}
\usepackage{chemformula}

\usepackage{afterpage}
\usepackage[markup=bfit]{changes}
\usepackage{xcolor}
\definechangesauthor[name={Bibiana Prinoth}, color=purple]{BP}
\definechangesauthor[name={Madeleine Burheim}, color=purple]{MB}
\definechangesauthor[name={Brian Thorsbro}, color=teal]{BT}
\definechangesauthor[name={Notes}, color=orange]{N}

\usepackage{booktabs}

\usepackage{color}
\usepackage{natbib,twoopt}
\usepackage[hyphenbreaks]{breakurl}
\usepackage[breaklinks]{hyperref}      
\bibpunct{(}{)}{;}{a}{}{,}             
\definecolor{cobalt}{rgb}{0.06, 0.2, 0.65}
\hypersetup{
  colorlinks,
  citecolor=cobalt,
  linkcolor=[rgb]{0.8, 0.2, 1.0},
  urlcolor=cobalt,
}
\makeatletter
  \newcommandtwoopt{\citeads}[3][][]{\href{http://adsabs.harvard.edu/abs/#3}%
    {\def\hyper@linkstart##1##2{}%
     \let\hyper@linkend\@empty\citealp[#1][#2]{#3}}}
  \newcommandtwoopt{\citepads}[3][][]{\href{http://adsabs.harvard.edu/abs/#3}%
    {\def\hyper@linkstart##1##2{}%
     \let\hyper@linkend\@empty\citep[#1][#2]{#3}}}
  \newcommandtwoopt{\citetads}[3][][]{\href{http://adsabs.harvard.edu/abs/#3}%
    {\def\hyper@linkstart##1##2{}%
     \let\hyper@linkend\@empty\citet[#1][#2]{#3}}}
  \newcommandtwoopt{\citeyearads}[3][][]%
    {\href{http://adsabs.harvard.edu/abs/#3}
    {\def\hyper@linkstart##1##2{}%
     \let\hyper@linkend\@empty\citeyear[#1][#2]{#3}}}
\makeatother

\newcommand{\jens}{WASP-121\,b\xspace}
\newcommand{\david}{WASP-76\,b\xspace}
\newcommand{\bibi}{WASP-189\,b\xspace}
\newcommand{\nic}{KELT-9\,b\xspace}
\newcommand{\nuria}{KELT-20\,b\xspace}

\newcommand{\elyar}{WASP-19\,b\xspace}

\newcommand{\kpvsys}{$K_p-V_{\rm sys}$\xspace}
\newcommand{\vorbtrue}{\num{200.7 \pm 4.9}}
\newcommand{\vsys}{\num{-24.452 \pm 0.012}}
\newcommand{\linn}{Boldt-Christmas et al. subm.\xspace}

\makeatletter
\renewcommand*\aa@pageof{, page \thepage{} of \pageref*{LastPage}}

\usepackage[separate-uncertainty = true,multi-part-units=single,group-separator={,}, group-digits = integer, group-four-digits = true,per-mode=reciprocal]{siunitx}

\newcommand{\numpm}[3]{$#1^{#2}_{#3}$}
\newcommand{\alldetections}{\ch{H}, \ch{Na}, \ch{Mg}, \ch{Ca}, \ch{Ca+}, \ch{Ti}, \ch{Ti+}, \ch{TiO}, \ch{V}, \ch{Cr}, \ch{Mn}, \ch{Fe}, \ch{Fe+}, \ch{Ni}, \ch{Sr}, \ch{Sr+}, and \ch{Ba+}\xspace}

\begin{document}
	
	\title{Time-resolved transmission spectroscopy of the ultra-hot Jupiter \bibi}
	
	\authorrunning{B. Prinoth et al. (2023)}
	
	\author{                                             
		B.\,Prinoth\inst{1}                              
		\and H.\,J.\,Hoeijmakers\inst{1}                 
		\and S.\,Pelletier\inst{2}                       
		\and D.\,Kitzmann\inst{3}                        
		\and B.\,M.\,Morris\inst{4}                      
		\and A.\,Seifahrt\inst{5}                        
		\and D.\,Kasper\inst{5}                          
		\and Heidi H. Korhonen\inst{6}          
		\and M.\,Burheim\inst{1,7}                       
		\and J.\,L.\,Bean\inst{5}                        
		\and B.\,Benneke\inst{2}                         
		\and N.\,W.\,Borsato\inst{1}                     
		\and M.\,Brady\inst{5}                           
		\and S.\,L.\,Grimm\inst{3,8}                     
		\and R.\,Luque\inst{5}                           
		\and J.\,Stürmer\inst{9}                         
		\and B.\,Thorsbro\inst{10,1}                     
	}
	
	\institute{
		Lund Observatory, Division of Astrophysics, Department of Physics, Lund University, Box 43, 221 00 Lund, Sweden 
		\and Department of Physics and Trottier Institute for Research on Exoplanets, Universit\'{e} de Montr\'{e}al, Montreal, QC, Canada 
		\and University of Bern, Physics Institute, Division of Space Research \& Planetary Sciences, Gesellschaftsstr. 6, 3012, Bern, Switzerland 
		\and Space Telescope Science Institute, Baltimore, MD 21218, USA 
		\and Department of Astronomy \& Astrophysics, The University of Chicago, Chicago, IL, USA 
		\and European Southern Observatory, Alonso de C\'ordova 3107, Vitacura, Regi\'on Metropolitana, Chile
		\and Material Sciences and Applied Mathematics, Malmö University, 205 06, Malmö, Sweden 
		\and ETH Zurich, Institute for Particle Physics and Astrophysics; Wolfgang-Pauli-Strasse 27, CH-8093 Z\"{u}rich, Switzerland 
		\and Landessternwarte, Zentrum für Astronomie der Universität Heidelberg, Königstuhl 12, 69117, Heidelberg, Germany 
		\and Observatoire de la C\^ote d'Azur, CNRS UMR 7293, BP4229, Laboratoire Lagrange, F-06304 Nice Cedex 4, France 
	}

	\offprints{Bibiana Prinoth, \\ \email{bibiana.prinoth@fysik.lu.se}}
	
	\date{}
	
	\abstract{Ultra-hot Jupiters are tidally locked with their host stars, dividing their atmospheres into a hot dayside and a colder nightside. As the planet moves through transit, different regions of the atmosphere rotate into view, revealing different chemical regimes. High-resolution spectrographs can observe asymmetries and velocity shifts and offer the possibility for time-resolved spectroscopy. The ultra-hot Jupiter \bibi has recently been found to possess a rich transmission spectrum with evidence for atmospheric dynamics and chemical inhomogeneity. In this study, we search for other atoms and molecules in the planet's transmission spectrum and investigate asymmetric signals. We analysed and combined eight transits of the ultra-hot Jupiter \bibi collected with the HARPS, HARPS-N, ESPRESSO, and MAROON-X high-resolution spectrographs. Using the cross-correlation technique, we searched for neutral and ionised atoms as well as oxides, and we compared the obtained signals to model predictions. We report significant detections for \alldetections. Of these,  \ch{Sr}, \ch{Sr+}, and \ch{Ba+} are detected for the first time in the transmission spectrum of \bibi. In addition, we robustly confirm the detection of titanium oxide based on observations with HARPS and HARPS-N using the follow-up observations performed with MAROON-X and ESPRESSO. By fitting the orbital traces of the detected species by means of time-resolved spectroscopy using a Bayesian framework, we inferred posterior distributions for orbital parameters as well as line shapes. Our results indicate that different species must originate from different regions of the atmosphere to be able to explain the observed time dependence of the signals. Throughout the course of the transit, most signal strengths are expected to increase due to the larger atmospheric scale height at the hotter trailing terminator. For some species, however, we instead observed that the signals weaken, either due to the ionisation of atoms and their ions or the dissociation of molecules on the dayside.
	}
	
	\maketitle
	
	\section{Introduction}
	\label{sec:Intro}
	
	\begin{figure*}
		\begin{minipage}{0.61\textwidth}
			\includegraphics[width=\linewidth]{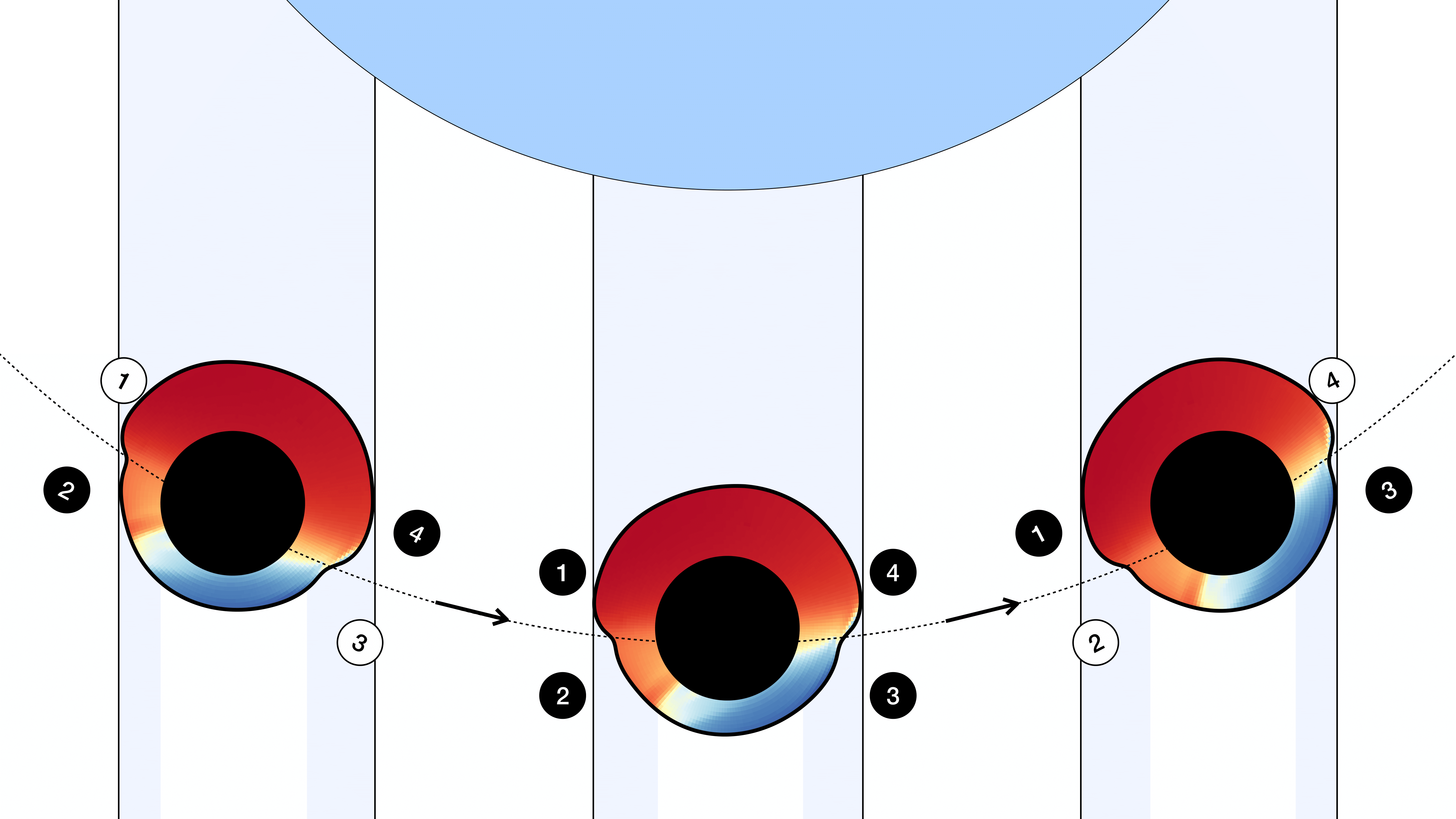}
			\label{fig:schematic}
		\end{minipage}
		\begin{minipage}{0.39\textwidth}
			\caption{Schematic of the contribution of different regions of the planetary atmosphere to the observed absorption signal over the course of the transit from left to right. As the planet rotates over the course of the transit, different regions become observable and dominate the contributions. Generally, the leading terminator (3, 4) is colder than the trailing terminator (1,2). The atmosphere at the trailing terminator is therefore expected to be more inflated (puffy) and generally more susceptible to transmission spectroscopy. The terminator regions that dominate the observed absorption signal are denoted with black circles. The colouring of the atmosphere is taken from \citet{lee_mantis_2022} to reflect the temperature of the atmosphere at a pressure of \SI{1}{mbar}. The shape of the atmosphere with an increased radius on the dayside is adopted from \citet{wardenier_all_2021}. We note that the schematic is not to scale. The changing viewing angle from the start to the centre of the transit for \bibi is $\alpha = \arcsin\left(\sqrt{1-b^2}\frac{R_\ast}{a}\right) \approx 11,33^\circ$, where $b$ is the impact parameter \citep[$b=$ \numpm{0.433}{+0.014}{-0.015}, ][]{deline_atmosphere_2022}, $R_\ast$ is the stellar radius, and $a$ is the semi-major axis.}
		\end{minipage}
	\end{figure*}

	Our understanding of planetary formation and evolution has been transformed by the discovery and characterisation of exoplanets over the past three decades. The hottest of these exoplanets are gas giant planets, often termed ultra-hot Jupiters, and they reside close to their host stars and typically have orbital periods of less than 10 days \citep{fortney_hot_2021}. As a result, ultra-hot Jupiters are subject to extreme variations in the temperature and chemistry of their atmosphere since they are thought to enter tidal confinement soon after their formation \citep{showman_atmospheric_2009,arras2010}.
	
	This tidal confinement, or tidal locking, of ultra-hot Jupiters divides their atmosphere into a permanently irradiated hot dayside and a cooler, permanently dark nightside. On the hot dayside, the high temperatures trigger processes such as thermal dissociation of most molecules, partial thermal ionisation of atomic species, and thermal inversions by metals and their oxides \citep{lothringer_extremely_2018, kitzmann_peculiar_2018, parmentier_thermal_2018, arcangeli_h-_2018}. On the colder nightside, we expect \ch{H2} recombination \citep{bell_increased_2018}, cloud formation by condensation, rain out, and cold traps of certain elements to the lower region of the planets, all of which are invisible to the distant observers \citep{spiegel_can_2009, helling_cloud_2021, komacek_patchy_2022, kesseli_atomic_2022, hoeijmakers_mantis_2022, gandhi_retrieval_2023, pelletier_vanadium_2023}. The close proximity of ultra-hot Jupiters to their host stars and their inflated atmospheres make them favourable targets for the study of atmospheric chemistry and dynamics using high-resolution transmission and emission spectroscopy. 
	Studies of ultra-hot Jupiters at optical wavelengths include, amongst others, the studies of \nic\citep{hoeijmakers_atomic_2018,hoeijmakers_spectral_2019,cauley_atmospheric_2019,pino_neutral_2020,pino_gaps_2022,borsato_mantis_2023}, \nuria\citep{casasayas-barris_na_2018,casasayas-barris_atmospheric_2019,hoeijmakers_high-resolution_2020,stangret_detection_2020}, \david\citep{ehrenreich_nightside_2020,tabernero_espresso_2021,seidel_into_2021,kesseli_atomic_2022,pelletier_vanadium_2023}, \jens\citep{hoeijmakers_hot_2020,hoeijmakers_mantis_2022,borsa_atmospheric_2021},  \bibi\citep{yan_temperature_2020,prinoth_titanium_2022}, WASP-33\,b \citep{cauley_time-resolved_2021,yan_detection_2021}, and \elyar\citep{sedaghati_spectral_2021}. In addition, the implementation of General Circulation Models using Monte Carlo simulations has significantly contributed to shaping our understanding of the atmospheric dynamics of ultra-hot Jupiters \citep{beltz_exploring_2021,wardenier_all_2021,beltz_magnetic_2022,lee_mantis_2022,savel_no_2022,beltz_magnetic_2023,savel_diagnosing_2023}.
	
	During a transit event, a tidally locked planet that is sufficiently close to its host star will rotate significantly during transit in such a way that different regions of the atmosphere become observable as the planet moves in front of the star on ingress, passes through the centre during mid-transit, and then leaves the line of sight on egress \citep{wardenier_all_2021}. In the case of some ultra-hot Jupiters, the rotation angle during transit is on the order of 20 to 40 degrees (e.g. \bibi: $\approx$\,23$^\circ$; \david: $\approx$\,28$^\circ$; \jens: $\approx$\,30$^\circ$; \nic: $\approx$\,36$^\circ$), causing different regions of the planetary atmosphere to contribute to the observed absorption spectrum. At the very beginning of the transit event, it is the hot side of the leading (morning) terminator that dominates the observed absorption, as the line of sight passes through more of the hotter regions at the leading terminator than at the trailing  (evening) terminator (see Fig.\,\ref{fig:schematic}). In the absence of any atmospheric dynamics, apart from co-rotation, this would result in a redshifted signal as the planet rotates counterclockwise. Similarly, towards the end of the transit, the absorption arises predominantly from the hot side at the trailing terminator, blueshifted by the planet's rotation. These effects lead to deviations in the apparent orbital velocity of the planet \citep{ehrenreich_nightside_2020, kesseli_confirmation_2021, wardenier_all_2021, kesseli_atomic_2022, prinoth_titanium_2022, pelletier_vanadium_2023}. Deviations in the observed systemic velocity of such planets suggest the presence of global dayside-to-nightside winds, while significant line broadening suggests the presence of a vertical outflow \citep[e.g.][]{seidel_into_2021,kesseli_atomic_2022,prinoth_titanium_2022}. Variations in the detected systemic and orbital velocities of different chemical species suggest that their absorption originates in different layers in the atmosphere \citep{prinoth_titanium_2022, maguire_high-resolution_2023}. These variations also suggest the presence of condensation and rain out on the nightside \citep{ehrenreich_nightside_2020,kesseli_atomic_2022}.

	Detections of absorbing species in the atmospheres of exoplanets are nowadays routinely made using the cross-correlation technique \citep{snellen_orbital_2010}. A key observable is obtained by stacking the time series of cross-correlation functions in the co-moving frame of the planet. In single exposures, the signal of the atmosphere may be too faint to be detected, whereas stacking in the rest frame of the planet increases the signal-to-noise ratio. While it is possible to optimise the exposure time to ensure a relatively high signal-to-noise ratio while avoiding smearing for a single transit event itself (\linn), a high signal-to-noise ratio can also be achieved by combining the data from multiple transit events. If the signal-to-noise ratio is sufficiently high, stacking in the rest frame of the planet is not necessary to see the atmospheric signature. In this case, the absorption signal can be studied using time-resolved spectroscopy to reveal empirical evidence for the presence of dynamics and chemical regimes probed at different times during the transit. While stacking in the rest frame of the planet effectively averages over the contributions from the different terminator regions, time-resolved spectroscopy allows the asymmetry in line shape and strength over the course of the transit to be studied and linked to the underlying chemistry and dynamics.

	Due to the strong thermal gradient caused by tidal locking,  different regions of the atmosphere may exhibit distinctly different chemical regimes. The hotspot may be significantly offset from the sub-stellar point \citep{showman_equatorial_2011}, making the trailing terminator hotter than the leading terminator, as it has a larger atmospheric scale height and deeper absorption lines when seen in transmission. In this region, the atmosphere may be hot enough to trigger the ionisation of atoms at certain altitudes.
	As an example, the ionisation of Ba to \ch{Ba+} requires less energy than the ionisation of lighter elements (e.g. Ca and Mg to \ch{Ca+} and \ch{Mg+}, respectively) within the same group. This is due to the fact that the effective nuclear charge experienced by the outer s-electron is constant within the group, whereas the orbital radius increases with the weight of the elements. Similarly, less energy is needed to ionise neutral atoms than singly-ionised atoms. The screening on the outer electron in an ionised atom is less efficient compared to the neutral atom, resulting in a larger effective nuclear charge and thus higher ionisation energy \citep{froese1997computational}. Similarly, species that are present on the nightside of the trailing terminator may condense out of the gas phase on the cooler nightside. Depending on the veracity of atmospheric dynamics, they may either get remixed on the dayside or be cold trapped \citep[e.g.][]{merritt_inventory_2021,pelletier_vanadium_2023}. Provided sufficient signal-to-noise, such effects may be observable using time-resolved spectroscopy. In this work, we present such observations of \bibi,  an ultra-hot Jupiter.\\
	
	The ultra-hot Jupiter \bibi is a gas\ giant planet orbiting the bright A-type star HD\,133112 \citep[V = 6.6\,mag;][]{anderson_wasp-189b_2018}. It is one of the most intensely irradiated planets known to date, with an equilibrium temperature of $\SI{2641 \pm 34}{K}$ \citep{anderson_wasp-189b_2018}, and it resides on a 2.7\,day long orbit that is most likely tidally circularised and synchronised. Recent observations by CHEOPS yielded constraining measurements of the planet radius, ephemeris, and a dayside temperature of $\SI{3400}{K}$ \citep{lendl_hot_2020}. Follow-up and refinement of the system parameters have been given by \cite{deline_atmosphere_2022}. The atmosphere of \bibi is inverted  \citep{yan_temperature_2020,yan_detection_2022}, based on the detection of emission by atomic iron and CO, which may in part be due to the presence of titanium oxide in its atmosphere \citep{prinoth_titanium_2022}. Observations with HARPS \citep{cosentino_harps-n_2012} and HARPS-N \citep{mayor_setting_2003} on the ESO 3.6m telescope on La Silla and the 3.58m telescope on the Telescopio Nazionale Galilei on La Palma, respectively, have revealed \ch{Fe}, \ch{Fe+}, and \ch{Ti} \citep{stangret_high-resolution_2022,prinoth_titanium_2022}, as well as \ch{Ti+}, \ch{V}, \ch{Mg}, \ch{Mn}, \ch{Cr}, and \ch{TiO} \citep{prinoth_titanium_2022}. Tentative detections include \ch{H} and \ch{Ca+} \citep{stangret_high-resolution_2022} as well as \ch{Cr+}, \ch{Sc+}, \ch{Na}, \ch{Ni}, and \ch{Ca} \citep{prinoth_titanium_2022}. These detections already showed evidence of dynamical effects (such as winds) through deviations in the apparent systemic and orbital velocities. Generally, the detected species appear blueshifted with respect to the expected systemic velocity due to prevailing day-to-night winds. \\

	In this work, we present the combined analysis of the high-resolution transmission spectrum of \bibi using eight transit events observed with the HARPS, HARPS-N, ESPRESSO, and MAROON-X spectrographs. We apply the cross-correlation technique \citep{snellen_orbital_2010} in order to search for atomic and ionised metals as well as molecules, and we fit the traces in the cross-correlation functions to determine spectral and orbital parameters using time-resolved spectroscopy. Section \ref{sec:DataObs} describes the observations and data reduction procedures. Section \ref{sec:methods} describes the telluric and preparatory corrections and further details the analysis process for the cross-correlation technique and trace fitting of this dataset, which presents the basis of our findings. In Section \ref{sec:results}, we present the results of the analysis and discuss their implications. Conclusions are presented in Section \ref{sec:Conclusions}.

	\section{Observations and data reduction}
	
	\begin{figure}[ht!]
		\centering
		\includegraphics[width=\linewidth]{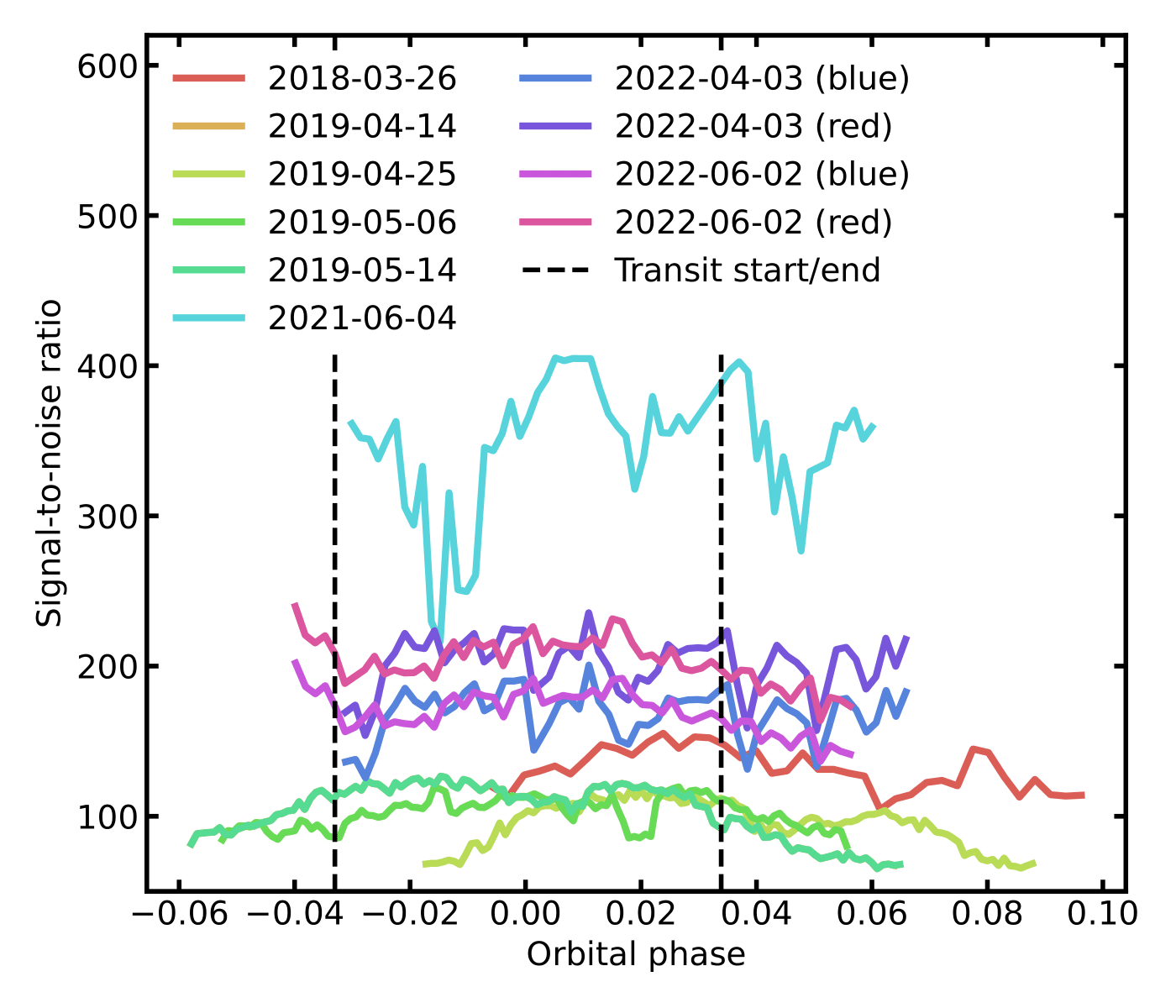}
		\caption{Signal-to-noise ratio versus orbital phase of each of the eight transit time series indicated in Table\,\ref{tab:observation_log}. The vertical dashed lines indicate the start and the end of the transit. The signal-to-noise ratio was calculated by averaging over all orders.}
		\label{fig:observation_coverage}
	\end{figure}
	
	\label{sec:DataObs}
	
	We observed one transit of the ultra-hot Jupiter \bibi using the ESPRESSO echelle spectrograph at the ESO Very Large Telescope in Chile \citep{pepe_espressovlt_2021} and two transits with the MAROON-X high-resolution optical spectrograph at the 8.1 m Gemini-North observatory in Hawaii \citep{seifahrt_maroon-x_2018,seifahrt_-sky_2020}. The transits were observed on 2021-06-04 with ESPRESSO (programme ID 107.22QF, PI: Prinoth) and on 2022-04-03 and 2022-06-02 with MAROON-X (programme ID GN-2022A-FT-208, PI: Pelletier).\\
	
	The ESPRESSO observations were performed in high-resolution mode with a spectral resolution of \(R = \frac{\lambda}{\Delta \lambda} = \num{140000} \) covering the wavelength range between $\num{380} - \SI{788}{nm}$. It is known that the readout speed of $500$ kpx\,s$^{-1}$ associated with the unbinned (1x1) readout mode introduces correlated noise at the bias level that is much larger than with the binned (2x1) readout mode. Therefore, we chose the binned readout mode, which is associated with a readout speed of $100$ kpx\,s$^{-1}$. Due to the initial poor weather conditions, the telescope dome was not opened until a few minutes before the start of the transit, causing the ingress to be partially missed. The observations of the in-transit spectra were obtained with UT-2. Due to a technical problem with UT-2 towards the end of the transit, the observations were continued with UT-1, causing the egress of the planetary transit to be lost. The out-of-transit observations were obtained with UT-1. As described in the ESPRESSO user manual, ESPRESSO spectra are influenced by two different interference patterns induced in the Coudé train optics.\footnote{\url{https://www.eso.org/sci/facilities/paranal/instruments/espresso/ESPRESSO_User_Manual_P108_v2.png}} It is known that approximately sinusoidal wiggles with periods of 30 and 1 Å and amplitudes of $\sim1\%$ and $\sim0.1\%$, respectively, become visible when spectra are recorded at different telescope positions and then divided by each other \citep{allart_wasp-127b_2020}. The high-frequency pattern (1 Å) is below the photon-noise level, while the low-frequency pattern was corrected as discussed in Section \ref{sec:prepcor}. Science observations were performed with fibre A on the target and fibre B on the sky.
	
	\begin{table*}[ht!]
		\caption{Overview of observations.}
		{\small \begin{center}
				\begin{tabular}{llllllll}
					\toprule
					Date & Instrument & PI, Prog. \# & $\#$ Spectra$^a$  & t$_{\rm exp}$ [s] $^{b}$  & Wavelength & Resolution $^{c}$ & Mirror\\
					& & & &   & coverage (wlc) [nm] $^{c}$ & & size [m] $^{c}$\\
					\midrule
					2018-03-26      & HARPS    & Hellier, 0100.C-0847          & 39 (15/24)        & 600                  & 383-693  & 120,000 & 3.6  \\ 
					2019-04-14      & HARPS    & Hoeijmakers, 0103.C-0472      & 126 (69/57)       & 200                  & 383-693  & 120,000 & 3.6  \\ 
					2019-04-25      & HARPS    & Hoeijmakers, 0103.C-0472      & 107 (52/55)       & 200                  & 383-693  & 120,000 & 3.6  \\
					2019-05-06      & HARPS-N  & Casasayas-Barris, CAT19A\_97  & 112 (69/43)       & 200                  & 383-690  & 115,000 & 3.6  \\
					2019-05-14  & HARPS    & Hoeijmakers, 0103.C-0472      & 122 (68/54)       & 200                  & 383-693  & 120,000 & 3.6  \\ 
					2021-06-04  & ESPRESSO & Prinoth, 107.22QF             & 55 (39/16)        & 300                  & 380-788  & 140,000 & 8.2 \\
					2022-04-03  & MAROON-X & Pelletier, GN-2022A-FT-208    & 57 (38/19)            & b: 200               & b: 490-720 & 80,000 & 8.1 \\
					& & &              & r: 160         & r: 640-920 &  & \\
					2022-06-02  & MAROON-X & Pelletier, GN-2022A-FT-208    & 57 (40/17)            & b: 200               & b: 490-720 & 80,000 & 8.1 \\
					& & &          & r: 160     & r: 640-920 & &  \\
					\bottomrule
				\end{tabular}
		\end{center}}
		\vspace{-1em}
		\textit{Note:} $^{(a)}$ In parentheses, in-transit and out-of-transit spectra, respectively.  $^{(b)}$ Exposure times differ for the blue and red arm (channel) of MAROON-X. $^{(c)}$ Adapted from \cite{kitzmann_mantis_2023} and references therein.
		\label{tab:observation_log}
	\end{table*}
	
	The observations performed with ESPRESSO were reduced using the ESPRESSO data reduction workflow (pipeline version 2.3.3) provided by ESO and executed via EsoReflex (version 3.13.5).\footnote{\url{https://www.eso.org/sci/software/pipelines/espresso/espresso-pipe-recipes.html}} The pipeline provides both one- and two-dimensional products with and without sky subtraction. The one-dimensional products contain a stitched, blaze-corrected spectrum, corrected for Earth's velocity around the barycentre of the Solar system (BERV) with both vacuum and air wavelength solutions. Within the scope of this paper, we generally denote the wavelength solution in air, unless mentioned otherwise. The two-dimensional products are saved per exposure. Each file contains all orders, including the wavelength solution in air.\\ 
	
	The observations with MAROON-X cover the wavelength range between \num{490} and \SI{900}{nm} with two arms (red and blue) at a spectral resolution of $\frac{\lambda}{\Delta\lambda} \approx\num{85000}$. The blue and red arms cover the wavelength ranges between \num{490} and \SI{720}{nm} and \num{640} and \SI{920}{nm} respectively. The observations cover the full transit, including baseline observations before and after the transit. 
	
	The MAROON-X observations were reduced using the standard pipeline \citep{seifahrt_-sky_2020}, which extracts the detector images into one-dimensional wavelength-calibrated spectra order by order for each of the exposures in the time series. For each of the channels (arms), the pipeline outputs a high-resolution spectral time series in the form of $N_{\rm exp} \times N_{\rm order} \times N_{\rm pixel}$ for each transit sequence. Respectively, $N_{\rm exp}$,  $ N_{\rm order}$, and  $N_{\rm pixel}$ denote the number of exposures, orders, and pixels for each detector. The pipeline provides the wavelength solution in vacuum, but we changed it to air for consistency between the observations.\\
	
	In addition to the newly obtained observations with ESPRESSO and MAROON-X, we used the transit observations of \bibi with HARPS-N \citep{cosentino_harps-n_2012} and HARPS \citep{mayor_setting_2003} previously published in \citet{anderson_wasp-189b_2018}, \citet{stangret_high-resolution_2022}, and \citet{prinoth_titanium_2022}. We used the reduced data as described in \citet{prinoth_titanium_2022}. A log of the time series observations used in this work is provided in Table \ref{tab:observation_log}, and an overview of the coverage of the observations is provided in Fig.\,\ref{fig:observation_coverage}.
	
	\section{Methods}
	\label{sec:methods}
	We follow the methodology as in \cite{prinoth_titanium_2022} and discuss deviations and necessary clarifications here in detail.  We further discuss the attempts to correct the problems with the interference patterns in the ESPRESSO data. 
	
	\subsection{Telluric correction}
	\begin{figure*}
		\centering
		\includegraphics[width=\textwidth]{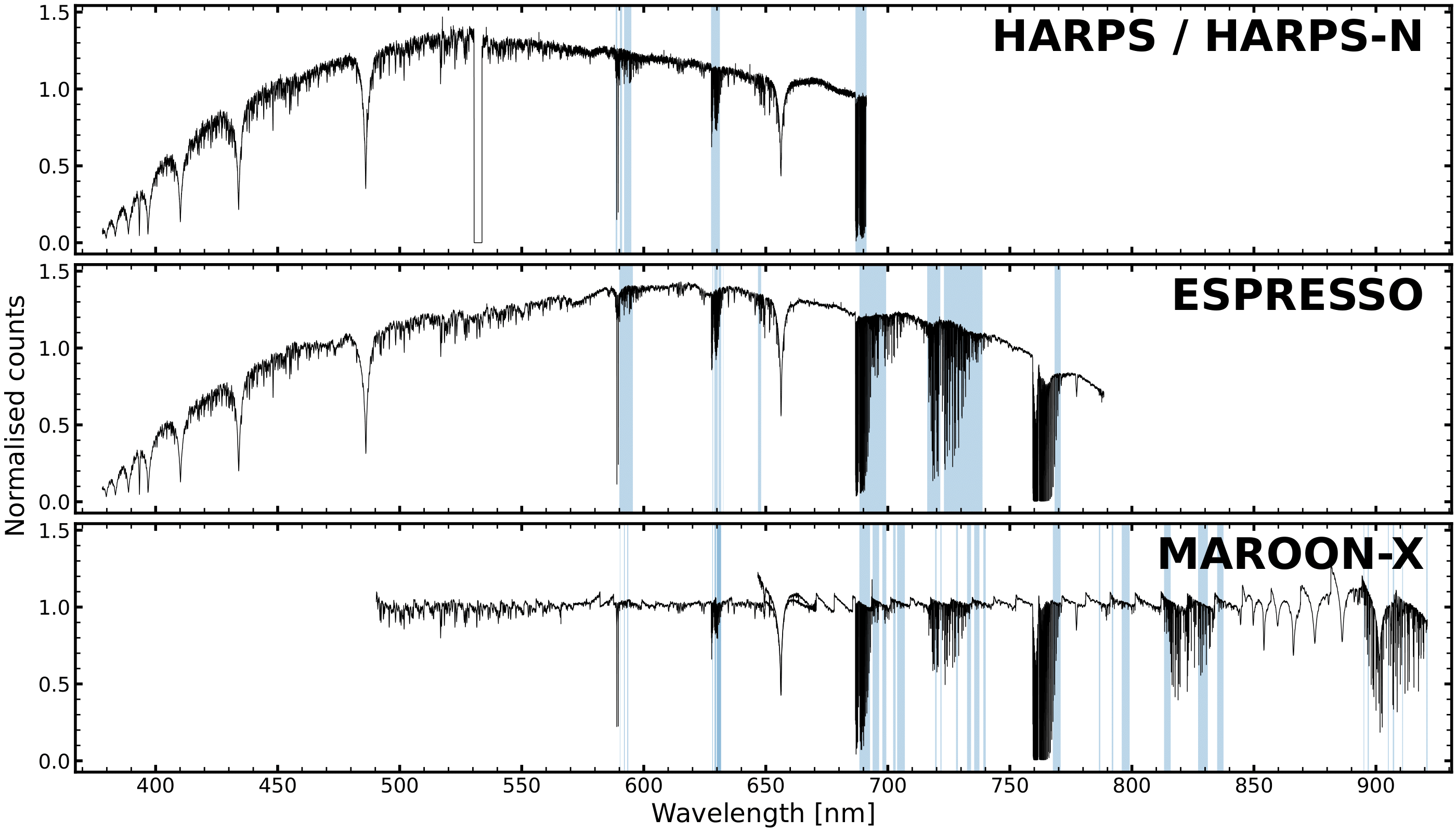}
		\caption{Extracted one-dimensional spectra and wavelength inclusion regions for telluric correction with \texttt{molecfit} for HARPS/ HARPS-N (top), ESPRESSO (middle), and MAROON-X (bottom). We note here that due to the simple de-blazing of the MAROON-X spectra, the continuum is not flat but instead has a slope. \texttt{Molecfit} takes care of this by fitting a higher-order polynomial for the continuum. Using these wavelength inclusion regions, we modelled the telluric profile for each exposure in each time series independently. MAROON-X observes two arms that overlap. The blue and red arms cover the wavelength ranges between \num{490} and \SI{720}{nm} and \num{640} and \SI{920}{nm}, respectively. We treated the two arms separately for our analysis due to different exposure times.}
		\label{fig:telluric_regions}
	\end{figure*}
	
	\label{sec:telluric_correction}
	Earth's atmospheric telluric lines were removed using \texttt{molecfit} \citep{smette_molecfit_2015,kausch_molecfit_2015}. It was applied to the one-dimensional spectra to create a model for the telluric transmission spectrum of the entire wavelength range. In the case of ESPRESSO, the reduction pipeline provides stitched and blaze-corrected one-dimensional spectra, and all header information in these files is constructed to be compatible with \texttt{molecfit}. For the MAROON-X observations, we manually stitched and de-blazed the spectrum in every order to create one-dimensional spectra for all the exposures. In addition, some of the keywords are different from ESO standards and need to be specified in the parameter file to run \texttt{molecfit}.\footnote{An example of the parameter file with comments and a script that puts the data in the right format to be read by \texttt{tayph} and later by \texttt{molecfit} in the framework of \texttt{tayph} can be found on GitHub at \url{https://github.com/bprinoth/MESOX.git}.}
	
	The telluric model computed with \texttt{molecfit} was run separately for each exposure in each time series. Regions with strong \ch{H2O} and \ch{O2} absorption bands were used to fit the model by selecting wavelength regions with telluric lines surrounded by a flat continuum with no stellar lines. The inclusion regions are shown in Fig. \ref{fig:telluric_regions}. The resulting telluric models were interpolated to the same wavelength grid as the individual spectral orders and then divided out to remove telluric absorption lines. Regions where the correction left visible residuals were later masked out manually. We note here that due to the simple de-blazing of the MAROON-X spectra, the continuum is not flat but has a slope instead.  This was taken care of by \texttt{molecfit}  by fitting a higher-order polynomial for the continuum. As shown in Fig. \ref{fig:telluric_regions}, there are not many telluric absorption bands in the blue arm of MAROON-X; notably only a strong \ch{O2} band around \SI{630}{\nm} ($\gamma$ band) and some weaker \ch{H2O} bands towards the end of the wavelength coverage of the arm can be seen. 
	
	\subsection{Preparatory corrections}
	\label{sec:prepcor}
	The individual spectra were Doppler-shifted to the rest frame of the host star by correcting for the Earth's velocity around the barycentre of the Solar system as well as the radial velocity of the star caused by the orbiting planet. This yielded a stellar spectrum with a constant velocity shift consistent with the systemic velocity of \vsys\,\si{\km\per\second}. Following \citet{hoeijmakers_high-resolution_2020}, we applied an order-by-order sigma clipping algorithm with a running median absolute deviation computed over sub-bands of the time series with a width of 40 pixels and rejected $5\sigma$-outliers. We further manually flagged spectral columns with visible systematic noise, where the telluric correction of deep lines presented residuals, most notably around known \ch{O2} and \ch{H2O} bands at redder wavelengths. For the red arm of the MAROON-X observations, outlier rejection as well as manual flagging affected, 
	at most, 25.56\% of the pixels due to strong telluric bands being entirely masked out. For the other observations, less than 2\% of the spectral pixels were affected. For the colour correction, we equalised the shape of the broadband continuum of each exposure in the time series, order by order, by fitting a polynomial of degree three. \\
	
	To account for the induced pattern in the ESPRESSO observations that became clearly visible when dividing the in-transit spectra by the out-of-transit spectra (see Fig. \ref{fig:ugly_pattern}), we introduced additional terms in the form of 
	
	\begin{equation}
		p_0 \sin\left(\frac{2 \pi x}{p_1}\right) + p_2 \cos\left(\frac{2 \pi x}{p_3} \right),
		\label{eq:sinusoid}
	\end{equation}
	
	where $p_i$ are the free parameters and $x$ the pixels per order. Dividing out the obtained fit removes any colour-dependent variations (polynomial) as well as the low-frequency pattern (sinusoidal). The sine and cosine terms in Eq. \eqref{eq:sinusoid} are only used on the ESPRESSO time series. The improvement by the inclusion of the sinusoidal fit was later shown to be insignificant, as the high-pass filter at the end of the cross-correlation workflow removes most residual variations (see Section\,\ref{sec:cleaning_steps}). A comparison of the cross-correlation results for \ch{Fe} with and without the sinusoidal component is shown in Fig.\,\ref{fig:cc_comp_fe}. We decided not to include this sinusoidal correction term when conducting our analysis.

	\subsection{Cross-correlation templates}
	For atoms and ions, we used the cross-correlation templates provided by \citet{kitzmann_mantis_2023}, and for TiO, we produced a separate cross-correlation template based on the Toto line list \citep[][states updated on 2021-08-25\footnote{ \url{https://www.exomol.com/data/molecules/TiO/48Ti-16O/Toto/}}]{mckemmish_exomol_2019}. For the cross-correlation templates, the atmosphere was assumed to be in isothermal and hydrostatic equilibrium, with temperatures of \SI{2500}{\kelvin} for TiO, \SI{3000}{\kelvin} for atoms, and \SI{4000}{\kelvin} for ions. The temperature of \SI{4000}{\kelvin} is motivated by the fact that most atoms are not significantly ionised below this temperature. For atoms, we used templates at a temperature just above the equilibrium temperature of \SI{2641}{\kelvin} \citep{anderson_wasp-189b_2018}. The templates and models were broadened to the full width at half maximum of \num{2.7}, \num{2.14}, and \SI{3.53}{\km\per\second}, matching the approximate line-spread functions of HARPS/ HARPS-N, ESPRESSO, and MAROON-X, respectively. A two-dimensional cross-correlation map was computed for each individual species template as described in \citet{prinoth_titanium_2022}. We used all the available templates in \citet{kitzmann_mantis_2023} with sufficient absorption lines at the considered temperatures, resulting in 90 cross-correlation templates available for this study, as indicated in Fig.\,\ref{fig:periodic_system}.

	\subsection{Cleaning steps}
	\label{sec:cleaning_steps}
	
	During transit the planet partially obscures areas of the rotating star disc, introducing residual spectral lines when performing differential transmission spectroscopy (Doppler shadow). In most cases, the Doppler shadow feature occurs at a different apparent radial velocity than the planet signature such that it can be removed without significantly affecting the planetary trace in the cross-correlation map \citep[for the case of the overlapping orbital trace and Doppler shadow of MASCARA-1\,b, see][]{casasayas-barris_transmission_2022}. Residing on a nearly polar orbit \citep{anderson_wasp-189b_2018}, \bibi introduces a Doppler shadow feature that is nearly constant in velocity space, appearing vertically in the cross-correlation map. It only overlaps with the planetary trace during egress, at the end of the transit. The overlapping region was excluded when co-adding the spectral time series to ensure that any residuals of the correction of the Doppler shadow do not affect the results. Because of the strong Fe absorption in the host star, we constructed empirical models for the Doppler shadow features of \ch{Fe} and \ch{Fe+} for atoms and ions respectively \citep[previously done in ][]{prinoth_titanium_2022} by fitting a function of the following form to each row in the two-dimensional cross-correlation map (i.e. each exposure $i$):
	
	\begin{align}
		\begin{split}
			f(v, i) &= A_1 A_{\rm poly} \exp\left(- \frac{(v - v_{\ast,i})^2}{2 (W_1  W_{\rm poly})^2}\right) \\
			&+ A_2 \exp\left(- \frac{(v - v_{\ast,i} - v_{\rm offset})^2 }{ 2W_2^2}\right) + C,
		\end{split}
		\label{eq:empirical_model}
	\end{align}
	
	where $A_i$ are the constant amplitudes of the two Gaussians, $W_i$ are the constant widths of the two Gaussians, and $C$ is a constant offset. Furthermore, $A_{\rm poly}$ and $W_{\rm poly}$ describe two polynomials of degree four to account for the centre-to-limb variations and the gravity darkening of the host star visible in the Doppler shadow features. These polynomials effectively allowed us to vary both the width and the amplitudes of the first Gaussian component such that the bright emission-like feature at the edges can also be removed successfully. The second Gaussian component corrects the wider absorption-like feature and additionally offers the possibility to be offset by a velocity shift of $v_{\rm offset} \sim \SI{46.55}{\km\per\second}$, which corresponds to half the projected rotational velocity of the star (see Table \ref{tab:parameters_XXX}).
	
	\begin{table}[ht!]
		\caption[]{Summary of stellar and planetary parameters of the WASP-189 system adopted in this study.}
		\small
		\begin{center}
			\def\arraystretch{1.25}
			\begin{tabular}{p{0.55\linewidth}p{0.3\linewidth}p{0.02\linewidth}}
				
				\toprule
				\multicolumn{3}{c}{Planetary parameters}  \\ 
				\midrule
				Planet radius ($R_{\rm p}$) [$R_{\rm Jup}$]                              & \numpm{1.600}{+0.017}{-0.016}                     & [3] \\
				Planet mass ($M_{\rm p}$) [$M_{\rm Jup}$]                                & \numpm{1.99}{+0.16}{-0.14}                        & [1] \\
				Eq. temperature ($T_{\rm eq}$) [$\si{\kelvin}$]                          & \num{2641+-31}                                    & [2] \\
				Density ($\rho$) [\si{\g\per\cm\cubed}]                                  & \numpm{0.6220}{+0.0769}{-0.0365}                  & [1] \\
				Surface gravity ($\log g_{\rm p}$)                                       & \numpm{18.8}{+2.1}{-1.8}                          & [1] \\      
				\midrule
				\multicolumn{3}{c}{Stellar parameters} \\ \midrule
				Star radius ($R_\ast$) [$R_{\odot}$]                                     & \num{2.36+-0.030}                                 & [1] \\
				Star mass ($M_\ast$) [$M_{\odot}$]                                       & \num{2.030+-0.066}                                & [1] \\
				Proj. rot. velocity ($v\sin{I_\ast}$) [\si{\km\per\second}]              & \num{93.1+-1.7}                                   & [1] \\
				Systemic velocity ($v_{\rm sys}$) [\si{\km\per\second}]                  & \num{-24.452+-0.012}                              & [2] \\
				\midrule
				\multicolumn{3}{c}{Orbital and transit parameters} \\ \midrule
				Transit centre time ($T_0$) [${\rm BJD}_{\rm TT}$ - 2450000]             & \numpm{8926.5416960}{+0.0000650}{-0.0000640}      & [1] \\
				Orbital semi-major axis ($a$) [au]                                       & \num{0.05053+-0.00098}                            & [1] \\
				Scaled semi-major axis ($a/R_\ast$)                                      & \numpm{4.600}{+0.031}{-0.025}                     & [1] \\
				Orbital inclination ($i$) [$^\circ$]                                     & \num{84.03+-0.14}                                 & [1] \\
				Projected orbital obliquity ($\lambda$) [$^\circ$]                       & \numpm{86.4}{+2.9}{-4.4}                          & [1] \\
				Eclipse duration ($T_{14}$) [h]                                          & \numpm{4.3336}{+0.0054}{-0.0058}                  & [1] \\
				Radius ratio ($R_p/R_\ast$)                                              & \numpm{0.07045}{+0.00013}{-0.00015}               & [1] \\
				RV semi-amplitude ($K$)  [\si{\km\per\second}]                           & \num{0.182+-0.013}                                & [1] \\
				Period ($P$) [d]                                                         & \num{2.7240330+-0.0000042}                        & [2] \\
				Eccentricity                                                             & 0.0                                               & [1] \\
				\midrule
				\multicolumn{3}{c}{Derived parameters$^\dagger$} \\
				\midrule 
				Orbital velocity $v_{\rm orb}$ [\si{\km\per\second}]                     &  \num{201.8+-3.9} & \\
				Proj. orbital velocity ($K_{\rm p} = v_{\rm orb}\sin i$)  [\si{\km\per\second}]& \vorbtrue                                    & \\          
				Approx. scale height ($H$) [\si{\km}]                                    & \numpm{579}{+65}{-56}                              & \\
				Transit depth of $H$ ($\Delta F / F$) {\scriptsize [$\times 10^{-5}$]}   & \numpm{4.94}{+0.57}{-0.49}                         & \\
				\bottomrule
			\end{tabular}
		\end{center}
		\label{tab:parameters_XXX}
		\textit{Note:} $^\dagger$ The projected orbital velocity is given by $K_{\rm p} = v_{\rm orb}\sin i = \frac{2\pi a \sin i}{P}$. The approximate scale height is given by $H = \frac{R T_{\rm eq}}{\mu g_{\rm p}}$, where $R$ is the gas constant and $\mu$ is the mean molecular weight of the atmosphere. In the case of \bibi, we calculated the mean molecular weight using \texttt{FastChem} \citep{stock_fastchem_2018,stock_fastchem_2022} and found $\mu = \SI{2.016}{\g\per\mol}$, which is in agreement with a mostly hydrogen and helium dominated atmosphere. The transit depth of a scale height is calculated using $\frac{\Delta F}{F} = \frac{A}{\pi R_\ast^2} = \frac{2HR_{\rm p} + H^2}{R_\ast^2}$. All uncertainties were calculated assuming Gaussian error propagation. References: [1] \cite{lendl_hot_2020}, [2] \cite{anderson_wasp-189b_2018}, [3] \cite{deline_atmosphere_2022}.
	\end{table}
	
	The local radial velocity $v_\ast$ in Eq. \eqref{eq:empirical_model} is given by the parameters of the system, such as the semi-major axis $a$, the stellar radius $R_\ast$, the phase of the transiting planet on its orbit $\phi$, the projected rotational velocity of the star $v\sin I_\ast$, the projected orbital obliquity $\lambda$, and the inclination of the system $i$ as stated in Table\,\ref{tab:parameters_XXX}. It takes the following form \citep{cegla_rossiter-mclaughlin_2016}
	
	\begin{equation}
		\mathbf{v_\ast(\phi)} = \frac{a}{R_\ast} \left(\sin 2\pi \mathbf{\phi} \cos \lambda + \cos 2\pi \mathbf{\phi} \cos i \sin\lambda\right) v\sin I_\ast.
	\end{equation}
	
	We scaled the empirical models for \ch{Fe} and \ch{Fe+} to fit the Doppler shadow of the individual species and divided out of the time-resolved two-dimensional cross-correlation map. We expected the projected rotational broadening of the tidally locked planetary atmosphere to be
	
	\begin{equation}
		\sigma_{\rm rot} = v_{\rm rot} \sin i = \frac{2 \pi R_p}{P} \sin i = 3.04\,\si{\km\per\second}
	\end{equation}
	
	using the parameters in Table\,\ref{tab:parameters_XXX}. While fitting the models, the planetary signature was protected by masking out the radial velocity range of the planet at each orbital phase, allowing for a width of \SI{16}{\km\per\second}of the mask to also ensure that broadened lines and potential winds of the order of 10 \si{\km\per\second} of the planetary atmosphere do not contribute to the fit.

	After the removal of the Doppler shadow, we performed a Gaussian high-pass filter with a width of \SI{100}{\km\per\second} that removes broadband structures present in the spectral direction, which also removes the pseudo-sinusoidal variations introduced by the reflections in the Coudé train visible in the ESPRESSO data. All the correlated structures and systematic noise constant in radial velocity that remained present after applying the Gaussian high-pass filter are not related to the planetary atmospheres and could thus be removed through detrending of the spectral channels individually following \citep{prinoth_titanium_2022,borsato_mantis_2023}. These structures are mainly caused by aliases between strong lines of the cross-correlation templates and the stellar absorption lines consecutively obscured by the transiting planet. They appear at constant velocities, thus vertically in the cross-correlation maps, and are again caused by the nearly polar alignment of the planetary orbit.
	
	\subsection{Model injection}
	\label{sec:model_inj}
	In order to predict the expected signal strength, we injected models into the data before performing the cross-correlation analysis. We followed the same procedure as in \citet{prinoth_titanium_2022} and used the same models at a temperature of \SI{2500}{\kelvin}, close to the planetary equilibrium temperature of \SI{2641}{\kelvin} \citep{anderson_wasp-189b_2018}, and \SI{3000}{\kelvin}, close to the dayside temperature \citep{lendl_hot_2020}, assuming an isothermal atmosphere in chemical and hydrostatic equilibrium of solar metallicity. The models include the continuum and accurate line depths and profiles. Prior to cross correlation, we injected model spectra into the observed spectra. When subtracting the true cross-correlation map (without injection) from the one with the injected signal, the Doppler shadow, aliases, and correlated noise were cancelled. This effectively left a residual signature signifying the predicted line depth and shape of the models.  
	
	\begin{figure*}[ht!]
		\centering
		\includegraphics[width=\textwidth]{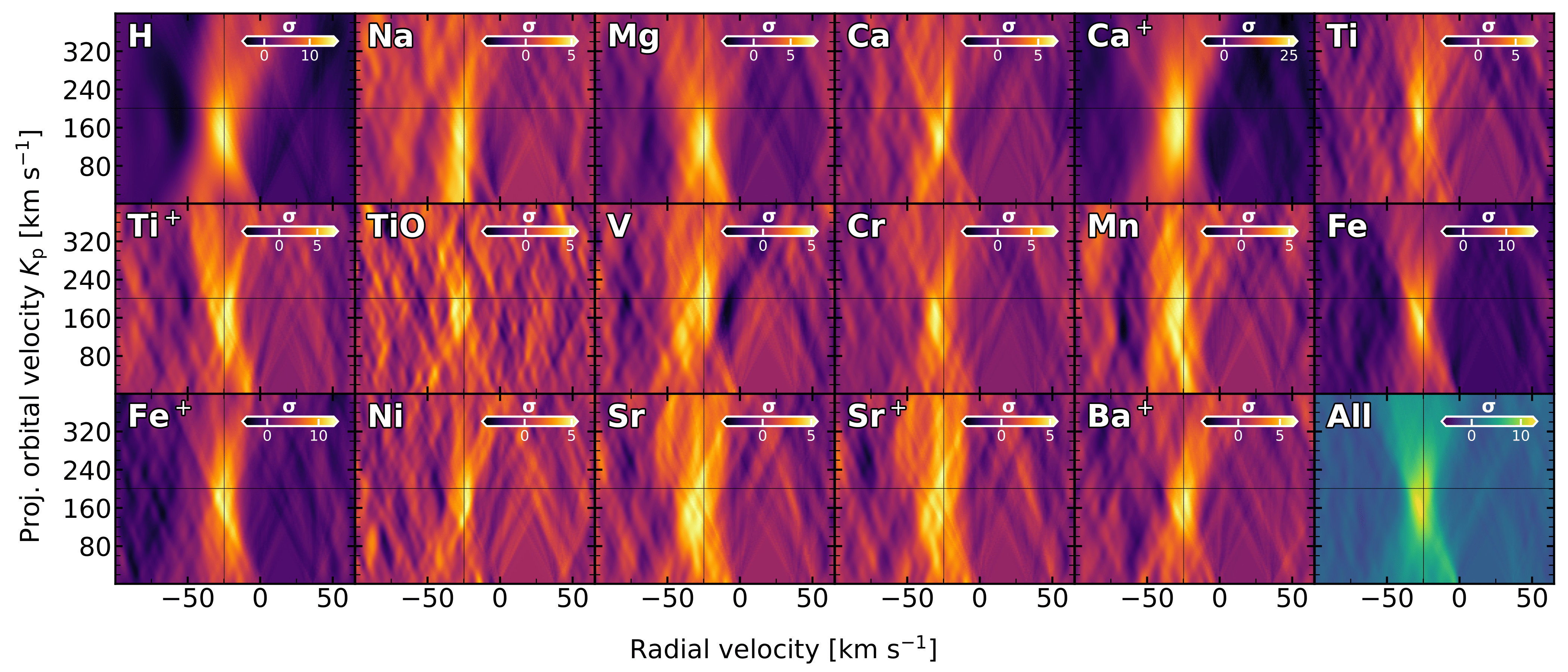}
		\caption{Cross-correlation results in the \kpvsys space for the species detected in \bibi's atmosphere during eight transit events with HARPS, HARPS-N, ESPRESSO, and MAROON-X combined. Each panel shows the significance of the detection for a given species, and the solid lines indicate the expected location of the signal if we assume a symmetric planet with a static atmosphere. Deviations from this location can be indicative of chemical asymmetries and dynamical effects. The lower-right panel shows the cross-correlation result in \kpvsys space for a template containing all species at a temperature of \SI{3000}{\kelvin}. As previously seen in \citet{prinoth_titanium_2022}, the atmospheric signal is located at a systemic and projected orbital velocity smaller than the true values, indicating a global dayside-to-nightside flow. }
		\label{fig:det}
	\end{figure*}
	
	\begin{figure*}[ht!]
		\centering
		\includegraphics[width=\linewidth]{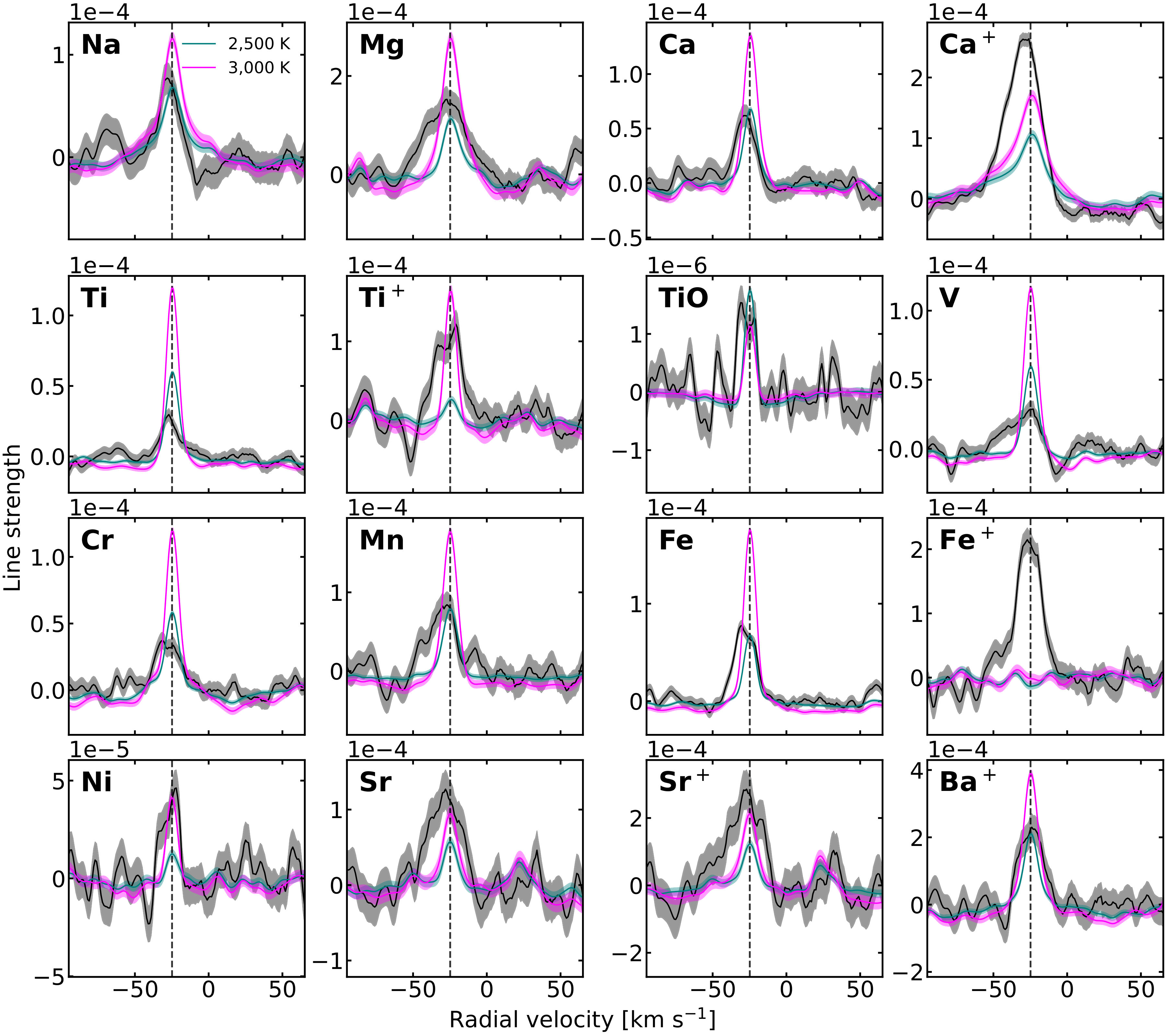}
		\caption{One-dimensional cross-correlation results extracted at the projected best-fit orbital velocity compared to the injected signal at $K_{\rm p} =$ \vorbtrue\,\si{\km\per\second}. The model predictions show the expected signal strength using models at \SI{2500}{\kelvin} (in teal) and \SI{3000}{\kelvin} (in magenta) in chemical and hydrostatic equilibrium of solar metallicity, previously used in \citet{prinoth_titanium_2022}. The shaded regions indicate the standard deviation away from the peak. Neutral atoms are generally well described by the model at \SI{2500}{\kelvin}. The dashed vertical line shows the expected systemic velocity of \vsys\,\si{\km\per\second}. The model prediction of TiO has been scaled by a factor of six to match the predicted signal strength in \citet{prinoth_titanium_2022} for the purpose of comparing the datasets from the instruments in Fig. \ref{fig:TiO_comparison}.}
		\label{fig:model_injection}
	\end{figure*}
	
	\subsection{Velocity-velocity maps}
	
	We then converted the resulting cleaned two-dimensional cross-correlation maps into velocity-velocity maps (\kpvsys diagrams) by shifting them towards the expected rest\ frame of the planet ($K_{\rm p}=$ \vorbtrue\,\si{\km\per\second}, $v_{\rm sys}=$ \vsys \si{\km\per\second}), assuming values of projected orbital velocity between $\num{0}$ and $\SI{400}{\km\per\second}$ in steps of \SI{1}{\km\per\second}. In each step, the exposures were weighted according to the mean flux of the corresponding spectra. The \kpvsys diagrams of the different observations were then combined by stacking and weighting based on the injected signals, where predicted (see Section \ref{sec:model_inj}), or based on the expected flux in the covered wavelength regions predicted by the template, if the species was not predicted by model injection. In the case of weighing by injection, this accounts for observing conditions, wavelength coverage, and signal-to-noise properties of the spectrographs. If weighted by the flux from the templates, we would only account for the different wavelength coverage. 
	
	\subsection{Time-resolved spectroscopy}
	\label{sec:time-resolved}
	The \kpvsys diagrams represent an average over the entire limb of the planet, though we know the signal will vary at each location on the limb, so it is challenging to identify the origin and nature of the contribution of the different regions of the atmosphere to the observed absorption signature \citep{wardenier_all_2021}. With the high signal-to-noise from the combined eight individual transit time series, it is possible to fit the planetary trace in the two-dimensional cross-correlation functions by means of time-resolved spectroscopy. As these cross-correlation functions are provided in the rest frame of the star, the planetary atmosphere traces a nearly linear part of the sinusoid of the radial velocity curve given as
	
	\begin{equation}
		\bm{v_{\rm planet}}(\bm{\phi}) =  v_{\rm orb}\sin\left(2\pi \bm{\phi}\right) \sin{i} + v_{\rm sys},
		\label{eq:planet_velocity}
	\end{equation}
	
	where $v_{\rm orb}$ is the orbital velocity of the planet, $\bm{\phi}$ is the orbital phase, $i$ is the inclination of the system, and $v_{\rm sys}$ is the systemic velocity. We modelled the two-dimensional absorption feature in transit as 
	
	\begin{align}
		\begin{split}
			{\rm \mathbf{CCF}} &= \mathbf{A}(\bm{\phi}, F_{\rm start}, F_{\rm end})\cdot \exp{\left(-\frac{(\bm{v} - \bm{v_{\rm planet,1}})^2}{2 \sigma^2}\right)} \cdot \bm{f} \\
			&+ \mathbf{A}(\bm{\phi}, F_{\rm start}, F_{\rm end})\cdot \exp{\left(-\frac{(\bm{v} - \bm{v_{\rm planet,2}})^2}{2 \sigma^2}\right)} \cdot \bm{f_{\rm inv}} + C,
		\end{split}
		\label{eq:ccf_gauss}
	\end{align}
	
	with $\bm{v_{\rm planet, i}}$ as defined in Eq.\,\eqref{eq:planet_velocity} for the two halves of the transit independently. We allowed the orbital velocity ${v_{\rm orb}}$ to vary between the two halves of the transit but required the systemic velocity $v_{\rm sys}$  to be the same for both halves, and thus it was fit at the centre of the transit. The terms $f$ and $f_{\rm inv}$ approximate two Heaviside step functions defined as
	
	\begin{align}
		\bm{f} &= \frac{1}{2}\tanh{k\bm{\phi}} + \frac{1}{2}\\
		\bm{f_{\rm inv}} &= -\frac{1}{2}\tanh{k\bm{\phi}} + \frac{1}{2}
	\end{align}
	
	which divide the transit into the first ($\bm{f}$) and second half ($\bm{f_{\rm inv}}$). We chose the $\tanh$ prescription of the Heaviside step function because our framework requires the model to be fully auto-differentiable to work with \texttt{NumPyro} and \texttt{Jax} \citep{jax2018github,bingham_pyro_2018,phan_composable_2019}. To account for the steep cut at the centre of the transit, we chose $k=1000$. The amplitude $\mathbf{A}(\bm{\phi}, F_{\rm start}, F_{\rm end})$ is given as a function of the phases of the transit, and the cross-correlation signal strength are indicated with $F_{\rm start}$ and $F_{\rm end}$ at the start and the end of the transit, respectively. This is given as   
	
	\begin{equation}
		\mathbf{A}(\bm{\phi}, F_{\rm start}, F_{\rm end}) = \frac{F_{\rm start} \phi_{\rm end} - F_{\rm end}\phi_{\rm start}}{\phi_{\rm end} - \phi_{\rm start}} - \frac{F_{\rm end} - F_{\rm start}}{\phi_{\rm end} - \phi_{\rm start}} \bm{\phi}.
	\end{equation}
	
	The ratio $\frac{F_{\rm start}}{F_{\rm end}}$ provides a first-order estimate for the increase ($> 1$) or decrease ($< 1$) in signal strength throughout the transit. The model in Eq. \eqref{eq:ccf_gauss} has the following free parameters and priors:
	
	\begin{itemize}
		\item The signal strength at the start of the transit $F_{\rm start} \sim \mathcal{U}(0, 1000)$ in parts per ten thousand, where $\phi_{\rm start} = -0.034 = 0.966$.
		\item The signal strength at the end of the transit $F_{\rm end} \sim \mathcal{U}(0, 1000)$ in parts per ten thousand, where $\phi_{\rm end} = 0.034$
		\item A constant offset $C \sim \mathcal{U}(-3,3)$ in ppm.
		\item The projected orbital velocities $K_{\textrm{p, i}} \sim \mathcal{U}(50,350)$ in \si{\km\per\second}, roughly centred on the expected projected orbital velocity of \vorbtrue\,\si{\km\per\second}.
		\item The systemic velocity $v_{\textrm{sys}}  \sim \mathcal{U}(-35,-15)$ in \si{\km\per\second}, roughly centred on the known systemic velocity of \vsys \si{\km\per\second}.
		\item The Gaussian line width $\sigma_{\rm w} \sim \mathcal{U}(1,14)$ in \si{\km\per\second}. The expected rotational broadening due to the planetary co-rotation is $\approx$\SI{3}{\km\per\second}, as given by $v_{\rm rot} = \frac{2 \pi R_p \sin i}{P}$ using the parameters stated in Table\,\ref{tab:parameters_XXX}. 
	\end{itemize}
	
	We sampled from these prior distributions and evaluated the likelihood in a Bayesian framework using a No-U-Turn Sampler \citep[see][for a review]{betancourt_convergence_2017}. Our implementation of this framework is based on \texttt{NumPyro} and \texttt{Jax} \citep{jax2018github,bingham_pyro_2018,phan_composable_2019}, as previously done for emission in \citet{hoeijmakers_mantis_2022}. We chose 200 warm-up samples and 600 samples over 20 chains running in parallel. When fitting the trace, we ignored the contributions from the region of the Doppler shadow entirely to avoid biasing the results by potential over- or under-correction of the stellar signal. \\
	
	Apart from a first-order estimate of the change in signal strength through the ratio $\frac{F_{\rm start}}{F_{\rm end}}$, the free parameters describe the shape and location of the trace, $v_{\textrm{sys}}$ describes the velocity of the atmosphere at mid-transit, and a deviation from the expected systemic velocity is indicative of the location of the species in the atmosphere and dynamics. We allowed different orbital velocities in the two halves of the transit \citep[see e.g.][for the shape of \ch{Fe} feature of WASP-76\,b]{ehrenreich_nightside_2020,kesseli_confirmation_2021,kesseli_atomic_2022}. Lastly, $C$ describes the continuum of the two-dimensional cross-correlation function.

	\section{Results and discussion}
	\label{sec:results}
	
	\subsection{Cross-correlation analysis}
	Figure\,\ref{fig:det} shows the 17 detected chemical species in the transmission spectrum of \bibi after combining eight transit time series obtained with HARPS, HARPS-N, ESPRESSO, and MAROON-X. We detected \alldetections. The detection significances including uncertainties are listed in Table\,\ref{tab:det_sign}.  While \citet{prinoth_titanium_2022} tentatively detected \ch{Cr+} and \ch{Sc+} at a velocity consistent with the projected orbital velocity of the planet, we were not able to confirm this detection when including three more transit observations. Apart from \ch{Cr+} and \ch{Sc+}, we confirm all detections and tentative detections in \citet{prinoth_titanium_2022} and \citet{stangret_high-resolution_2022}. Additionally, we detected \ch{Sr}, \ch{Sr+}, and \ch{Ba+}, which have not been detected in the atmosphere of this planet previously. We report detections for both \ch{Sr} and \ch{Sr+}, but we note that the signal shapes are irregular (more broadened) in comparison to the rest of the detections. This requires further investigation.

	\subsection{Model comparison}
	Figure\,\ref{fig:model_injection} shows the expected significance when injecting the models introduced in Section \ref{sec:model_inj}. The absorption lines of most neutral atoms are broadly consistent with a model at a temperature of \SI{2500}{\kelvin} in local thermodynamic equilibrium (LTE) and hydrostatic and chemical equilibrium. This agreement was already observed in \cite{prinoth_titanium_2022}. As previously observed in WASP-121\,b \citep{hoeijmakers_high-resolution_2020}, the signal strength for \ch{Ni} is higher than predicted by any of the two models. The strong absorptions of \ch{Fe+} and \ch{Ca+} are inconsistent with this class of models, where \ch{Fe+} is not predicted to be observable, while \ch{Ca+} is predicted to be weaker, even at a model at \SI{3000}{\kelvin}. Departures from model predictions, especially in the case of ions, are likely reflective of the unknown temperature-pressure profiles at high altitudes, the possible breakdown of hydrostatic equilibrium for the deepest lines, and non-LTE effects \citep[see also][]{prinoth_titanium_2022, ehrenreich_nightside_2020, kesseli_atomic_2022, savel_no_2022}. The broad shape of the \ch{Ca+} feature is consistent with absorption at high altitudes as part of an extended outflow \citep{yan_ionized_2019,seidel_into_2021,tabernero_espresso_2021,maguire_high-resolution_2023, pelletier_vanadium_2023}.
	We found that \ch{Ba+} is consistent with the model prediction at \SI{2500}{\kelvin} and \ch{Sr+} at \SI{3000}{\kelvin}, while \ch{Ti+} requires a model at a higher temperature to match the observed signal. In general, the predictions for ionised species show that this class of models does not reliably predict the absorption depths observed in ultra-hot Jupiters. 
	
	\subsection{Robust confirmation of TiO}
	
	\begin{figure*}
		\centering
		\includegraphics[width=
		\linewidth]{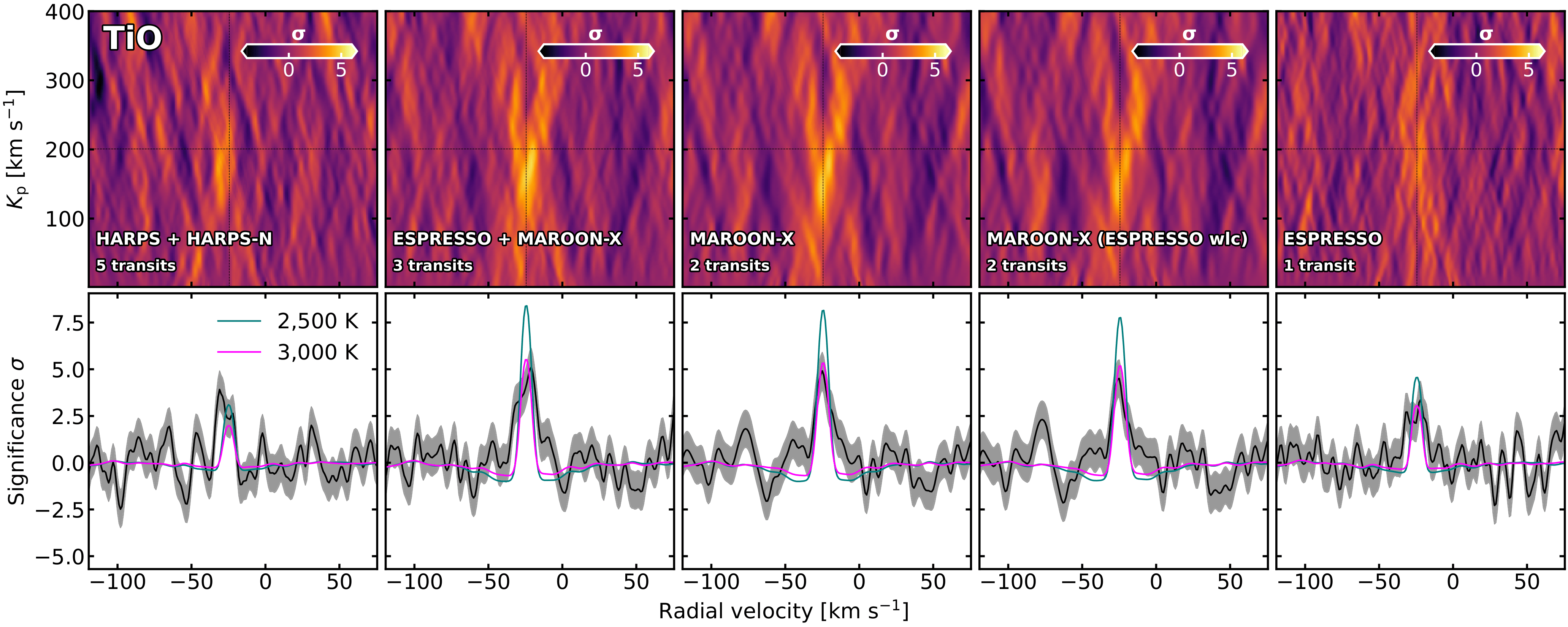}
		\caption{Cross-correlation results in \kpvsys space for \ch{TiO} in \bibi's atmosphere during eight transit events with HARPS \& HARPS-N, ESPRESSO, and MAROON-X. From left to right: MAROON-X alone, MAROON-X restricted to ESPRESSO wavelength coverage (wlc), and ESPRESSO alone (see Table\,\ref{tab:observation_log}). The dashed lines indicate the expected location of the signal if we assume a symmetric planet with a static atmosphere. The model predictions show the expected signal strength using models at \SI{2500}{\kelvin} (in teal) and \SI{3000}{\kelvin} (in magenta) in chemical and hydrostatic equilibrium of solar metallicity, previously used in \citet{prinoth_titanium_2022}, and have been scaled by a factor of six for clarity to show that the model discrepancy is equivalent in the data from the three different instruments despite their different wavelength coverage. The predicted signals are a measure of the sensitivity of the dataset and show that we expect MAROON-X to have a stronger signal. We note that discrepancies with \citet{prinoth_titanium_2022} stem from a different weighting algorithm based on the predicted signal strength (through model injection) applied here. This weighting algorithm favours the signal of better nights, increasing the signal strength overall. The results show that the signal is recovered in the new datasets alone as well but with a lower relative flux compared with expectations from the model injection.\vspace{0.5cm}}
		\label{fig:TiO_comparison}
	\end{figure*}
	
	\begin{figure*}
		\centering
		\includegraphics[width=0.8\linewidth]{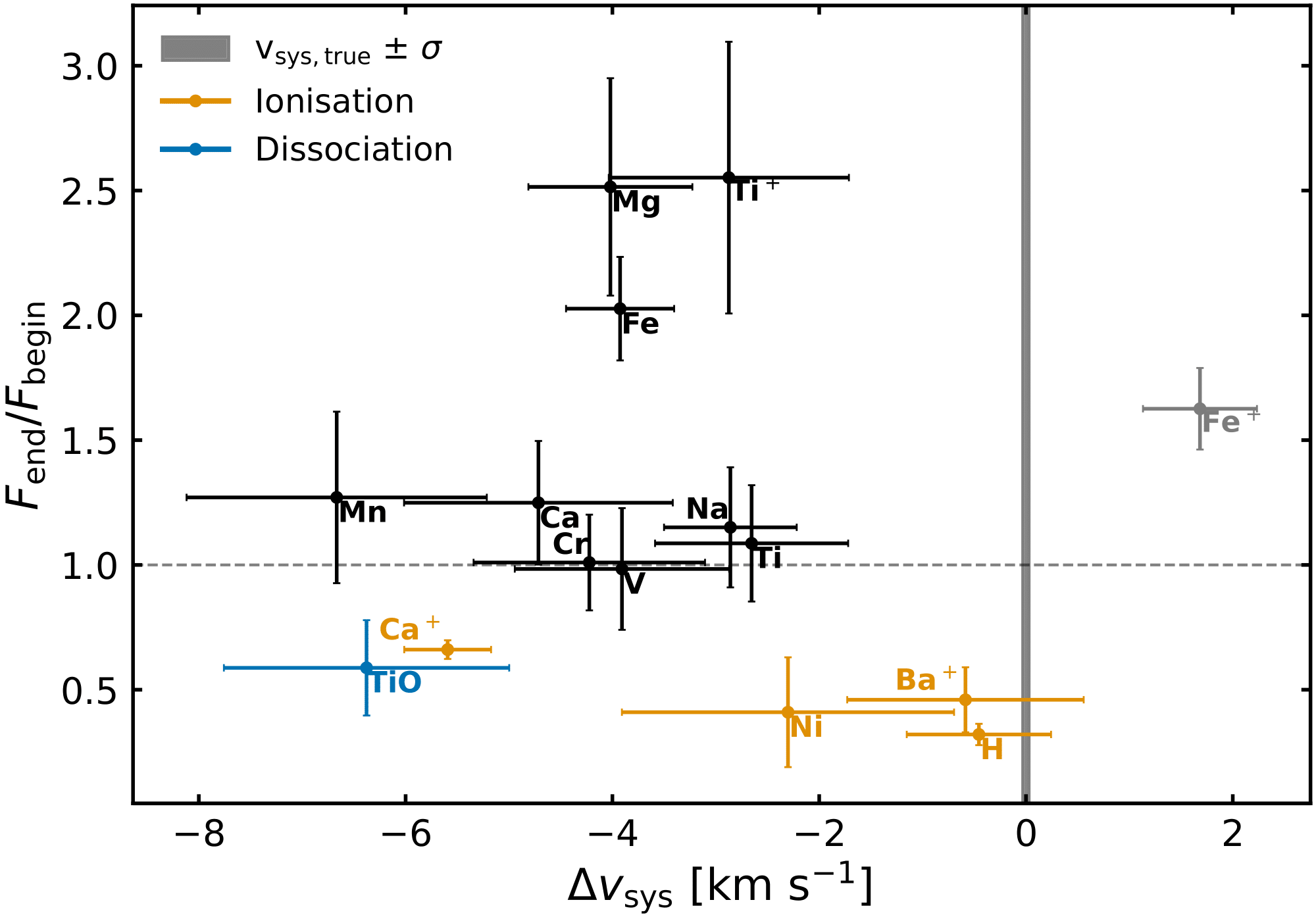}
		\caption{Change in signal strength between the start and the end of the transit as a function of the observed deviation from the expected systemic velocity. The error bars denote 1-$\sigma$ uncertainties calculated via Gaussian error propagation. The grey vertical region indicates the true systemic velocity range (\vsys \si{\km\per\second}). We note that \ch{Sr} and \ch{Sr+} are not included in this figure due to large uncertainties. We also note that \ch{Fe+} is an outlier (see Fig.\,\ref{fig:kinked_feature_fail}).  The hotter atmosphere rotating into view (see Fig.\,\ref{fig:schematic}) causes most signal strengths to increase over the course of transit.  However, further ionisation (\ch{Ca+}, \ch{Ni}, \ch{Ba+}, and \ch{H}) or dissociation (\ch{TiO}) on the dayside can also cause some signals to decrease over time.}
		\label{fig:deltavs}
	\end{figure*}
	
	The combination of the two MAROON-X and the ESPRESSO observations robustly confirms the detection for \ch{TiO} in the transmission spectrum of \bibi. Combining the data of all eight transit time series, we detected \ch{TiO} at 6.65 $\sigma$. The observed significance is in agreement with the model at \SI{2500}{\kelvin} if scaled by a factor of six \citep[needed to bring the model prediction in agreement with ][]{prinoth_titanium_2022}. 
	Despite the additional noise caused by interference patterns in the ESPRESSO data in the form of internal reflections in the optical path, it is possible to detect and thus confirm \ch{TiO} (see Fig.\,\ref{fig:TiO_comparison}). The detection of TiO in the MAROON-X and ESPRESSO data combined is mostly driven by the MAROON-X data.
	
	\begin{figure*}
		\centering
		\includegraphics[width=\linewidth]{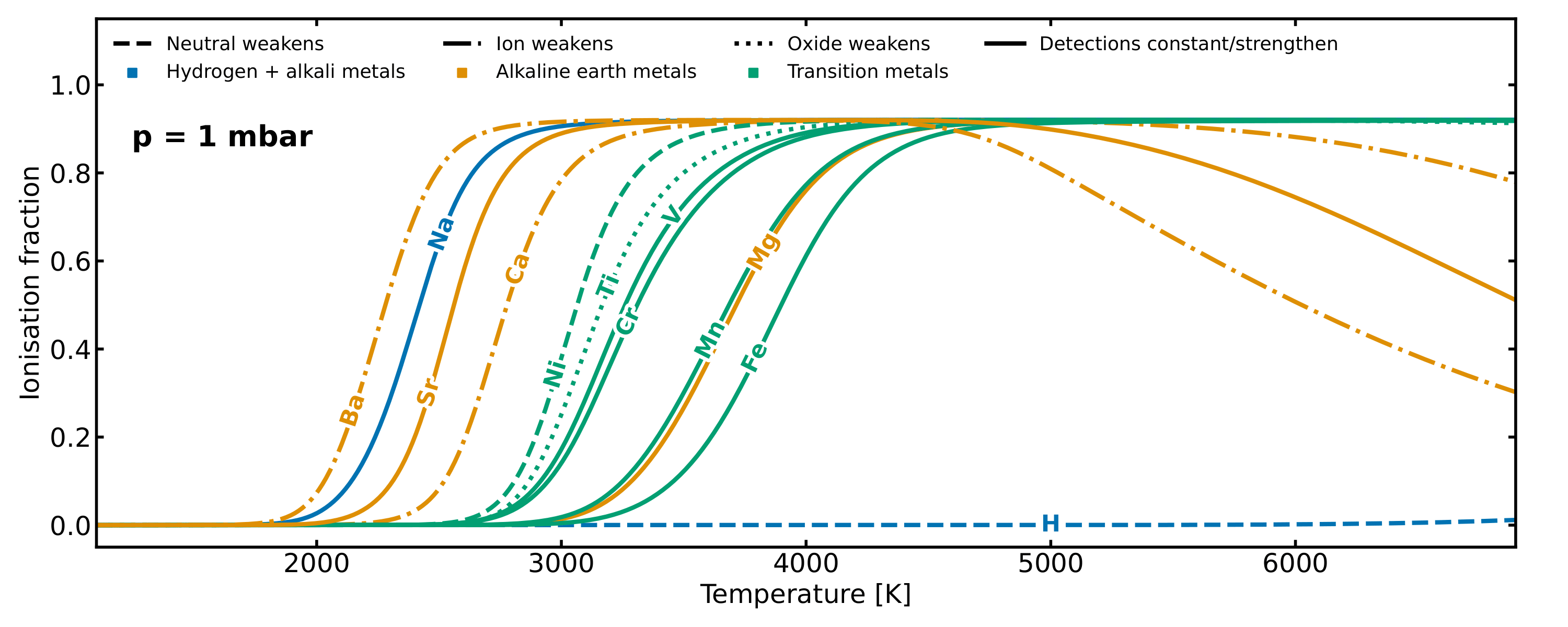}
		\includegraphics[width=\linewidth]{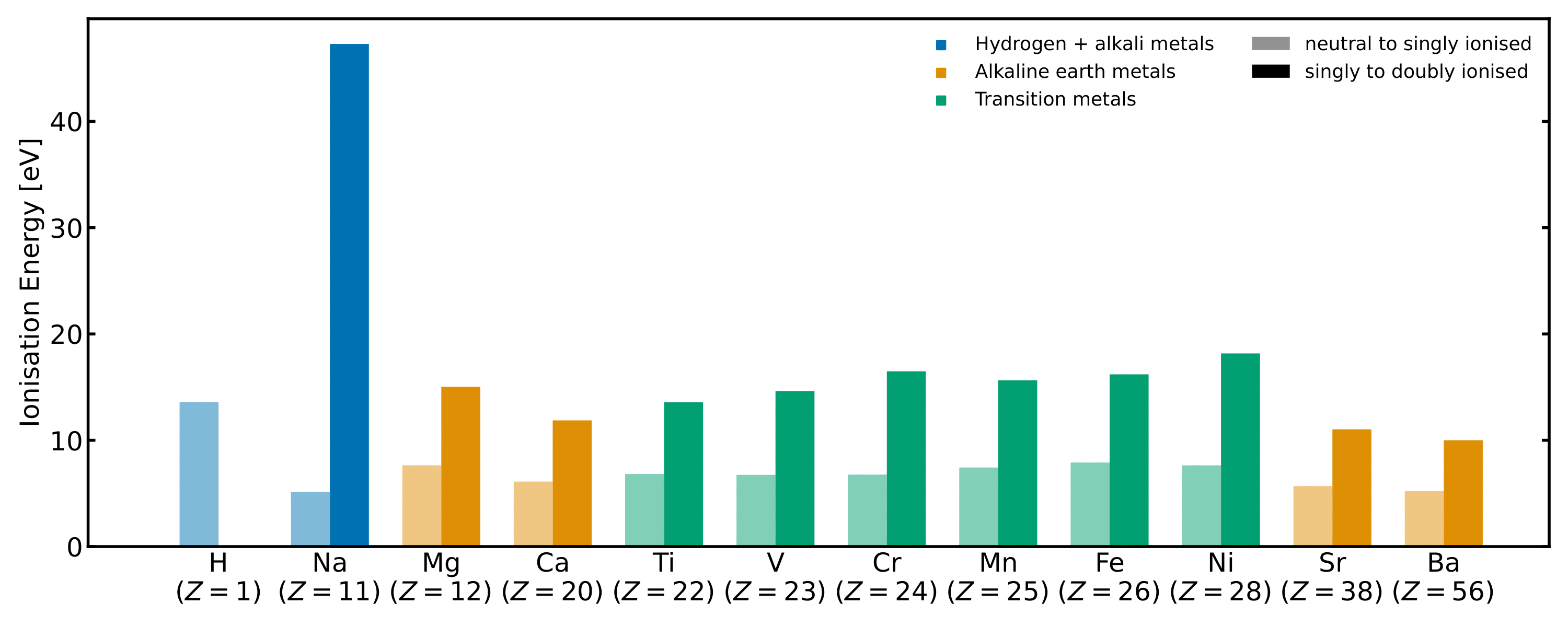}
		\caption{Ionisation fraction and energies for the atoms detected in the atmosphere of \bibi. The ionisation curves (top panel) were computed using \texttt{FastChem}, assuming chemical equilibrium at a pressure of 1 mbar and solar metallicity. Group 1 elements (hydrogen + alkali metals) are plotted in blue, and group 2 elements (alkaline earth metals) are plotted in orange. Transition metals are plotted in teal. The dashed, dash-dotted, and dotted lines indicate that the detected signal strength of the neutral atom, ionised atom, and oxide weakens over the course of the transit, respectively. Solid lines indicate that the signal strengths of the detections either strengthen over the course of the transit or remain constant, and thus behave as expected. The ionisation curves move towards lower temperatures with decreasing pressure (increasing altitude). Additionally, the fractions decrease at high temperatures for alkaline earth metals due to a second ionisation. The ionisation energies (bottom panel) are from the NIST database \citep{NIST_ASD} which predict that Na, Ba, Ca, \ch{Ba+}, and Sr are all easily ionised. Regarding H, it may significantly ionise at very high altitudes. Because of the constant effective nuclear charge within groups 1 and 2, the increasing orbital radius causes the ionisation energy to decrease for higher atomic numbers $Z$ (within the group). Within the same period (row), the ionisation energy increases due to the stronger attractive force of the nucleus. Both trends (within period and group) are generally not true for transition metals due to electron shells not necessarily being filled (and ionised) consecutively \citep{froese1997computational}.}
		\label{fig:ionisation}
	\end{figure*}
	
	\subsection{Time-resolved spectroscopy} 

	Combining the eight transit time series enabled the investigation of the time dependence of signals through time-resolved spectroscopy, as described in Section\,\ref{sec:time-resolved}. Figures\,\ref{fig:trace_Na}--\ref{fig:Ba_p_trace} show the results of the orbital trace fitting, and the best-fit parameters are summarised in Table\,\ref{tab:det_sign}. \\
	
	We observed changes in the signal strength of nine of the detected species, as shown in Fig.\,\ref{fig:deltavs}. The atmosphere was expected to be colder and less inflated at the leading terminator (east side), whereas both temperature and scale height increased when at the trailing terminator (west side), see Fig.\,\ref{fig:schematic}. The observations towards the end of the transit are therefore dominated by the light that has passed through the trailing terminator, so an increase in signal strength is consistent with the increased scale height there. With the exception of \ch{H}, \ch{Ba+}, \ch{Ca+}, \ch{Ni}, and \ch{TiO}, all detected species either showed an increase in signal strength over the course of the transit or appeared to be constant. \\
	
	A decrease in signal strength may be caused by a loss of the absorbing species towards the hot trailing terminator. We propose that this can be explained through either ionisation or dissociation on the dayside. Figure\,\ref{fig:ionisation} shows the ionisation temperatures and energies for the detected atomic species for the first and second ionisation states. The ionisation curves were computed using \texttt{FastChem} \citep{stock_fastchem_2018, stock_fastchem_2022}, assuming chemical equilibrium.  The ionisation energies were taken from the NIST database \citep{NIST_ASD}. The temperature on the dayside is likely high enough to thermally ionise species with relatively lower ionisation energies. Particularly notable are \ch{Ba+} and \ch{Ni}, which show a decrease in signal strength as hotter regions rotate into view, where they are more susceptible to ionisation (secondary ionisation in the case of \ch{Ba+}). We infer that the absorption signal of \ch{Ba+} predominantly originates from the hot region of the leading terminator (region 4 in Fig.\,\ref{fig:schematic}), which rotates out of view over the course of the transit. The same argument may also hold for \ch{Ni}, as it is expected to ionise on the dayside through to \ch{Ni+}, and for \ch{H} and \ch{Ca+}, as they are easily ionised to \ch{H+} and \ch{Ca^{2+}}, respectively, at higher temperatures. As \ch{Ni+} predominantly absorbs at wavelengths below the coverage of ESPRESSO (bluest wavelength range of the spectrographs used), a search for it is currently limited by the blue cut-off of the spectrograph. The signal strength of \ch{TiO} decreases over the course of the transit, while \ch{Ti+} increases and \ch{Ti} appears to be constant. We hypothesise that \ch{TiO} dissociates on the dayside \citep{cont_detection_2021}, while \ch{Ti} partly ionises to \ch{Ti+}, leaving the signal strength of \ch{Ti} roughly constant. At the same time, molecular dissociation at high temperatures may also act to replenish the neutral reservoir. For example, \ch{Ca} is one of the most refractory elements \citep{lodders_solar_2003}, though it also has a relatively low ionisation temperature. This means that the signal strength of neutral calcium on the opposite limbs results from a balance between depletion by ionisation and creation from molecular dissociation. More advanced chemical models including condensates may be needed to self-consistently explain these time-dependent signals.\\
	
	\citet{azevedo_silva_detection_2022} recently reported the detection of \ch{Ba+} in the transmission spectrum of \jens \citep[$T_{\rm eq}= \SI{2358(52)}{\kelvin}$,][]{delrez_wasp-121_2016}) and \david \citep[$T_{\rm eq}= \SI{2160(40)}{\kelvin}$,][] {west_three_2016}, and \citet{borsato_mantis_2023} detected it in the transmission spectrum of \nic \citep[$T_{\rm eq}= \SI{3921(180)}{\kelvin}$,][]{borsa_gaps_2019}. We fit the atmospheric trace of \ch{Ba+} obtained by \citet{borsato_mantis_2023} to show that this signal also weakens over the course of the transit (see Fig.\,\ref{fig:trace_Ba_nic}).\\

\begin{figure}
		\centering
		\includegraphics[width=\linewidth]{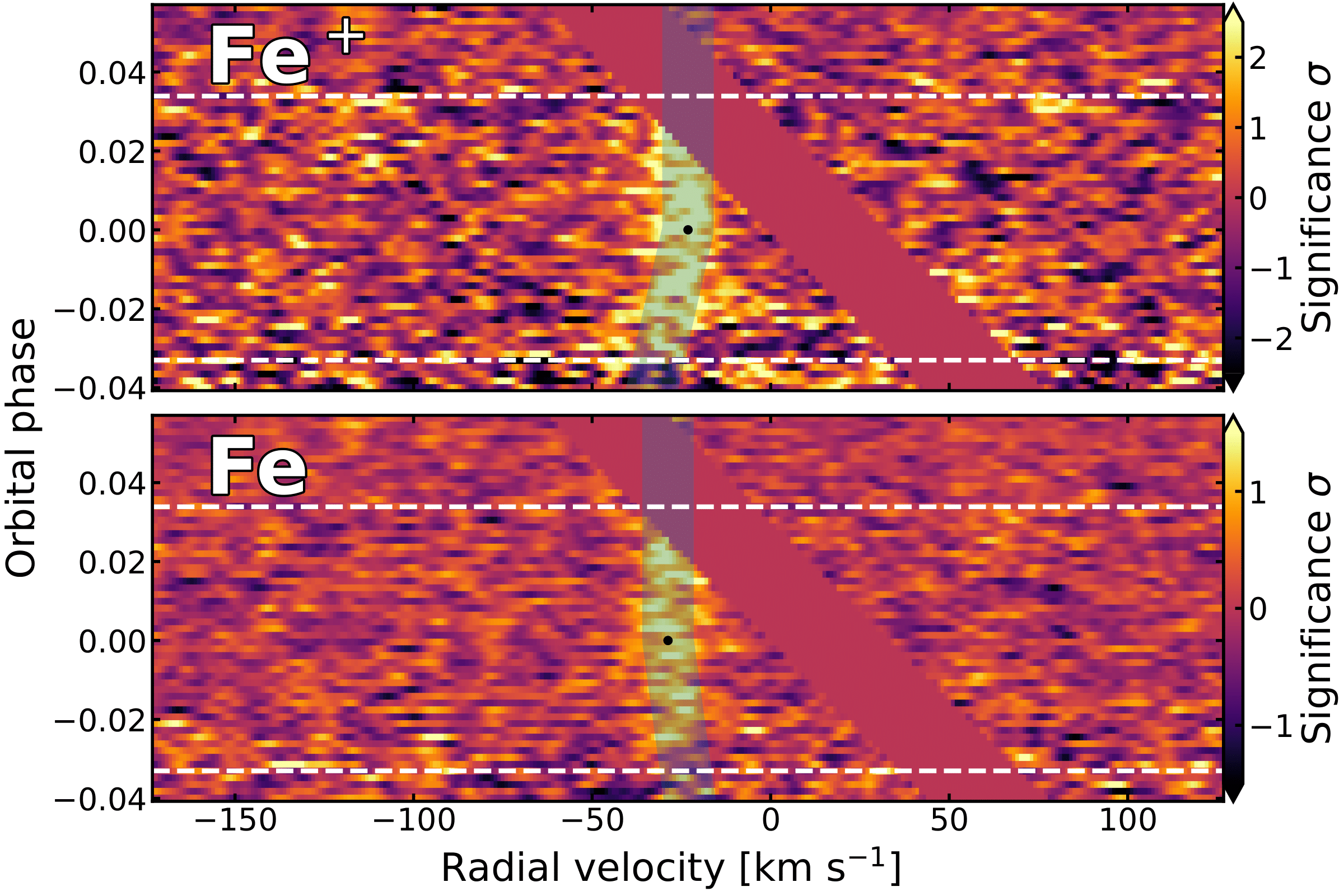}
		\caption{Two-dimensional cross-correlation function of \ch{Fe+} for all eight transit time series combined. The cross-correlation function is shifted to the projected orbital velocity found for the second part of the transit, $v_{\rm orb, 2}$ to illustrate the earlier change of orbital velocity. We purposely limited the colour bar in order to enhance the absorption feature that helps make the misfit visible. The dot indicates the centre of the transit. The blue shaded region shows the best-fit result.}
		\label{fig:kinked_feature_fail}
	\end{figure}

	
	Figure\,\ref{fig:deltavs} shows the change in signal strength between the start and the end of the transit as a function of the observed deviation from the true systemic velocity of \vsys \si{\km\per\s} for each individual species. With the exception of \ch{Fe+}, all of the detected species are blueshifted with respect to the true systemic velocity of \vsys\,\si{\km\per\s}, indicative of dayside-to-nightside winds and flows at high altitudes. The orbital fit for \ch{Fe+} suggests that the measured systemic velocity at the centre of the transit is redshifted with respect to the true systemic velocity. This is likely not astrophysical. Our fit works well for species that do have a roughly constant orbital velocity ($v_{\rm orb,1} \sim v_{\rm orb,2}$) or where the change between the two velocities is at the centre of the transit. Figure\,\ref{fig:kinked_feature_fail} shows that for \ch{Fe+}, the change in orbital velocity should happen earlier. The trace of the \ch{Fe+} feature is thus inconsistent with the best-fit model, as the model does not allow for change at other times than mid-transit. \\
	
	\subsection{Future prospects}
	The strongest signals are exhibited by \ch{Ca+} and \ch{Ba+}, inviting further analysis using narrow-band spectroscopy \citep[e.g.][]{seidel_detection_2023}. Notably, narrow-band spectroscopy has been successfully used in investigations of the \ch{Na} doublet \citep[e.g.][]{casasayas-barris_na_2018,seidel_hot_2019,seidel_into_2021,seidel_detection_2023} and will be further discussed in Prinoth et al. (in prep.). As for Na, \ch{Ca+}, and \ch{Ba+}, they are expected to behave similarly, with only a few very strong lines in their transmission spectra.
	
	\section{Conclusions}
	\label{sec:Conclusions}
	
	In this study, we searched for 90 different species, including atoms, ions, and various oxides, in the transmission spectrum of \bibi using the cross-correlation technique on eight transit time series with HARPS, HARPS-N, ESPRESSO, and MAROON-X. We detected \alldetections at high significance. We further confirmed the detection for \ch{TiO} and added novel detections for \ch{Sr}, \ch{Sr+}, and \ch{Ba+}. We note that \ch{Sr} and \ch{Sr+} have a deviating line shape that requires further investigation.\\
	
	For each of the detected species, we fit the orbital trace by means of a Bayesian framework to infer posterior distributions for orbital parameters as well as the line depth and shape as a function of time. For all species except \ch{Fe+}, the fit resulted in a blueshifted signal with respect to the true systemic velocity. We determined that the apparent redshift for \ch{Fe+} is due to the change in orbital velocity that is offset from the centre of the transit. While the signal strength of the majority of the species increases over the course of the transit, as expected when probing the hotter trailing terminator (increased scale height), the signal strength of \ch{Ni}, \ch{Ba+}, \ch{Ca+}, \ch{H}, and \ch{TiO} decreases instead. Throughout the course of the transit, as hotter regions of the atmosphere rotate into view, the signal of certain species weakens. We posit that this is likely caused by ionisation through incoming stellar radiation on the dayside that causes the atmosphere to heat up significantly (\ch{Ni}, \ch{Ba+}, \ch{Ca+}, \ch{H}). For \ch{TiO}, the weakened signal is likely similarly caused by dissociation on the hot dayside. \\
	
	Time-resolved spectroscopy offers a new possibility to study the atmospheres of hot gas giant exoplanets. By combining several transit observations, the signal-to-noise ratio becomes sufficiently high to study the absorption of the atmosphere resolved in the orbital phase, and thus, it is no longer necessary to average the observations in the rest frame of the planet in order to detect absorbing species. Instead of averaging over the contributions of the two terminators, time-resolved spectroscopy enables studying the change in signal strength and dynamics over the course of the transit as different regions of the tidally locked atmosphere rotate into view. It is especially useful for short-orbit planets with plenty of available observations, preferably those with different spectrographs so that a wide wavelength range can be covered (e.g. \david, \jens, \nic, and \nuria.) \\
	
	Ultimately, these observations need to be interpreted using models that accurately describe dissociation and ionisation processes (whether thermal or driven by the radiation field) as well as nightside condensation. Explaining this three-dimensional complexity of ultra-hot Jupiters will likely require models that simulate global circulation as well as radiative transfer in three dimensions.
	
	\begin{table*}
		\caption{Overview of detections.}
		{\scriptsize
			\begin{tabular}{llllllllll}
				\toprule
				& $F_{\rm start}$ [$\times 10^{-4}$] & $F_{\rm end}$ [$\times 10^{-4}$] & $A$ [$\times 10^{-4}$]& $\sigma_{\rm lw}$ [$\times 10^{-4}$] & $v_{\rm sys}$ [km s$^{-1}$] & $v_{\rm orb,1}$ [km s$^{-1}$] & $v_{\rm orb,2}$ [km s$^{-1}$] & $C$ [$\times 10^{-6}$] & $\sigma$ \\
				\midrule
				H       &  2604.71 $\pm$ 98.53 & 835.42 $\pm$ 106.81 & 1720.06 $\pm$ 72.66 & 11.19 $\pm$ 0.37 & -25.29 $\pm$ 0.70 & 197 $\pm$ 12.96 & 157.49 $\pm$ 6.33 & 0.14 $\pm$ 1.24 & 23.67 \\
				Na      & 169.76 $\pm$ 27.27 & 195.33 $\pm$ 26.16 & 182.54 $\pm$ 18.89 & 6.68 $\pm$ 0.62 & -27.69 $\pm$ 0.64 & 145.07 $\pm$ 7.76 & 145.07 $\pm$ 7.76 & 0.16 $\pm$ 0.33 & 9.66 \\
				Mg      & 257.48 $\pm$ 40.41 & 647.45 $\pm$ 47.43 & 452.47 $\pm$ 31.15 & 10.86 $\pm$ 0.64 & -28.85 $\pm$ 0.79 & 109.18 $\pm$ 7.79 & 203.58 $\pm$ 13.05 & 0.22 $\pm$ 0.53 & 14.52 \\
				Ca      & 116.96 $\pm$ 17.99 & 146.09 $\pm$ 18.25 & 131.52 $\pm$ 12.82 & 7.43 $\pm$ 0.69 & -29.55 $\pm$ 1.3 & 148.59 $\pm$ 13.57 & 178.49 $\pm$ 15.63 & 0.02 $\pm$ 0.22 & 10.26 \\
				Ca$^+$  & 785.0 $\pm$ 22.61 & 518.62 $\pm$ 25.39 & 651.81 $\pm$ 17.0 & 10.58 $\pm$ 0.23 & -30.42 $\pm$ 0.42 & 182.27 $\pm$ 4.04 & 184.11 $\pm$ 7.85 & -0.34 $\pm$ 0.33 & 38.34 \\
				Ti      & 64.92 $\pm$ 8.88 & 70.53 $\pm$ 11.66 & 67.73 $\pm$ 7.33 & 6.63 $\pm$ 0.57 & -27.48 $\pm$ 0.93 & 194.52 $\pm$ 12.37 & 215.96 $\pm$ 15.41 & -0.02 $\pm$ 0.1 & 9.24 \\
				Ti$^+$  & 182.29 $\pm$ 34.38 & 465.17 $\pm$ 46.38 & 323.73 $\pm$ 28.87 & 8.43 $\pm$ 0.64 & -27.7 $\pm$ 1.16 & 146.65 $\pm$ 15.4 & 225.35 $\pm$ 16.45 & 0.05 $\pm$ 0.43 & 11.21 \\
				TiO     & 4.17 $\pm$ 0.74 & 2.45 $\pm$ 0.67 & 3.31 $\pm$ 0.5 & 4.53 $\pm$ 0.64 & -31.21 $\pm$ 1.38 & 145.75 $\pm$ 12.61 & 208.8 $\pm$ 16.44 & -0.0 $\pm$ 0.01 & 6.65 \\
				V & 70.8 $\pm$ 11.49 & 69.66 $\pm$ 13.05 & 70.23 $\pm$ 8.69 & 9.21 $\pm$ 0.82 & -28.74 $\pm$ 1.04 & 207.5 $\pm$ 10.17 & 207.5 $\pm$ 10.17 & -0.06 $\pm$ 0.14 & 8.08 \\
				Cr      & 96.51 $\pm$ 12.84 & 97.45 $\pm$ 13.22 & 96.98 $\pm$ 9.22 & 7.52 $\pm$ 0.7 & -29.05 $\pm$ 1.12 & 165.65 $\pm$ 12.05 & 177.93 $\pm$ 14.58 & 0.05 $\pm$ 0.15 & 10.52 \\
				Mn      & 180.66 $\pm$ 32.36 & 229.53 $\pm$ 46.52 & 205.1 $\pm$ 28.33 & 8.27 $\pm$ 1.02 & -31.5 $\pm$ 1.45 & 159.33 $\pm$ 17.61 & 251.63 $\pm$ 28.15 & 0.11 $\pm$ 0.41 & 7.24 \\
				Fe      & 122.19 $\pm$ 11.02 & 247.64 $\pm$ 12.0 & 184.91 $\pm$ 8.14 & 7.62 $\pm$ 0.29 & -28.76 $\pm$ 0.52 & 171.65 $\pm$ 6.92 & 195.83 $\pm$ 6.67 & 0.06 $\pm$ 0.13 & 22.7 \\
				Fe$^+$  & 454.94 $\pm$ 35.28 & 739.59 $\pm$ 47.25 & 597.26 $\pm$ 29.49 & 8.26 $\pm$ 0.38 & -23.15 $\pm$ 0.55 & 218.0 $\pm$ 7.09 & 177.55 $\pm$ 7.73 & 0.14 $\pm$ 0.44 & 20.26 \\
				Ni      & 136.63 $\pm$ 22.26 & 56.0 $\pm$ 28.64 & 96.32 $\pm$ 18.14 & 5.56 $\pm$ 0.65 & -27.13 $\pm$ 1.61 & 180.04 $\pm$ 13.59 & 222.06 $\pm$ 49.0 & -0.13 $\pm$ 0.23 & 5.31 \\
				Sr      & 39.52 $\pm$ 31.54 & 640.32 $\pm$ 55.83 & 339.92 $\pm$ 32.06 & 12.49 $\pm$ 0.8 & -24.35 $\pm$ 1.04 & 233.02 $\pm$ 11.85 & 233.02 $\pm$ 11.85 & 0.18 $\pm$ 0.61 & 10.6 \\
				Sr$^+$  & 172.74 $\pm$ 86.49 & 1432.13 $\pm$ 136.44 & 802.44 $\pm$ 80.77 & 11.32 $\pm$ 0.82 & -24.12 $\pm$ 0.94 & 229.06 $\pm$ 10.13 & 229.06 $\pm$ 10.13 & 0.27 $\pm$ 1.24 & 9.93 \\
				Ba$^+$  & 853.22 $\pm$ 89.18 & 392.28 $\pm$ 103.35 & 622.75 $\pm$ 68.25 & 7.16 $\pm$ 0.53 & -25.42 $\pm$ 1.14 & 175.01 $\pm$ 9.89 & 182.69 $\pm$ 18.58 & -0.12 $\pm$ 0.97 & 9.12 \\
				\bottomrule
		\end{tabular}}\\
		\textit{Note:} The average signal strength $A$ is the mean of the signal strength at the start and end of the transit, $F_{\rm start}$ and $F_{\rm end}$. We note that for \ch{Na}, \ch{V}, \ch{Sr}, and \ch{Sr+}, the orbital velocity was forced to be the same for both regions of the transit because the posterior distribution reached the boundaries of the feasible priors. The detection significance $\sigma$ was calculated from the average signal strength $A$ and its uncertainty.
		\label{tab:det_sign}
	\end{table*}
	
	\begin{acknowledgements}
		B.P., H.J.H., and N.W.B acknowledge partial financial support from The Fund of the Walter Gyllenberg Foundation. S.P. and B.B. acknowledge financial support from the Natural Sciences and Engineering Research Council (NSERC) of Canada and the Fond de Recherche Québécois-Nature et Technologie (FRQNT; Québec). M.Br. acknowledges support from the National Science Foundation Graduate Research Fellowship under Grant No. DGE 1746045. R.L. acknowledges funding from the University of La Laguna through the Margarita Salas Fellowship from the Spanish Ministry of Universities ref. UNI/551/2021-May 26, and under the EU Next Generation funds. B.T.\ acknowledges the financial support from the Wenner-Gren Foundation (WGF2022-0041). M.Bu. acknowledges support from the Swedish Research Council VR through grant no. 2016-04185, and the Crafoord Foundation. We thank E.K.H Lee for providing us with the temperature map of \bibi at a pressure of 1 mbar published in \citet{lee_mantis_2022}. We thank Per Jönsson for his support in understanding the screening effect. We thank Romain Allart for discussions about the interference patterns in the ESPRESSO data. This study makes use of \texttt{astropy} \citep{astropy:2013,astropy:2018,astropy:2022}, \texttt{NumPyro} \citep{bingham_pyro_2018}, \texttt{
			matplotlib-label-lines} \citep{cadiou_matplotlib_2022}, \texttt{corner.py} \citep{foreman-mackey_cornerpy_2016}, \texttt{ArviZ} \citep{kumar_arviz_2019} and \texttt{JAX} \citep{jax2018github}. This work is based in part on observations collected at the European Southern Observatory under ESO programme 107.22QF. This research has made use of the services of the ESO Science Archive Facility.
		This work is based in part on data obtained at the international Gemini Observatory, a programme of NSF’s NOIRLab. The international Gemini Observatory at NOIRLab is managed by the Association of Universities for Research in Astronomy (AURA) under a cooperative agreement with the National Science Foundation on behalf of the Gemini partnership: the National Science Foundation (United States), the National Research Council (Canada), Agencia Nacional de Investigación y Desarrollo (Chile), Ministerio de Ciencia, Tecnología e Innovación (Argentina), Ministério da Ciência, Tecnologia, Inovações e Comunicações (Brazil), and Korea Astronomy and Space Science Institute (Republic of Korea). The MAROON-X team acknowledges funding from the David and Lucile Packard Foundation, the Heising-Simons Foundation, the Gordon and Betty Moore Foundation, the Gemini Observatory, the NSF (award number 2108465), and NASA (grant number 80NSSC22K0117). We thank the staff of the Gemini Observatory for their assistance with the commissioning and operation of the instrument.
		This work was enabled by observations made from the Gemini North telescope, located within the Maunakea Science Reserve and adjacent to the summit of Maunakea. We are grateful for the privilege of observing the Universe from a place that is unique in both its astronomical quality and its cultural significance. 
	\end{acknowledgements}


\newpage
\appendix

\onecolumn

\section{Interference pattern}

\begin{figure*}[ht!]
     \centering
     \includegraphics[width=\linewidth]{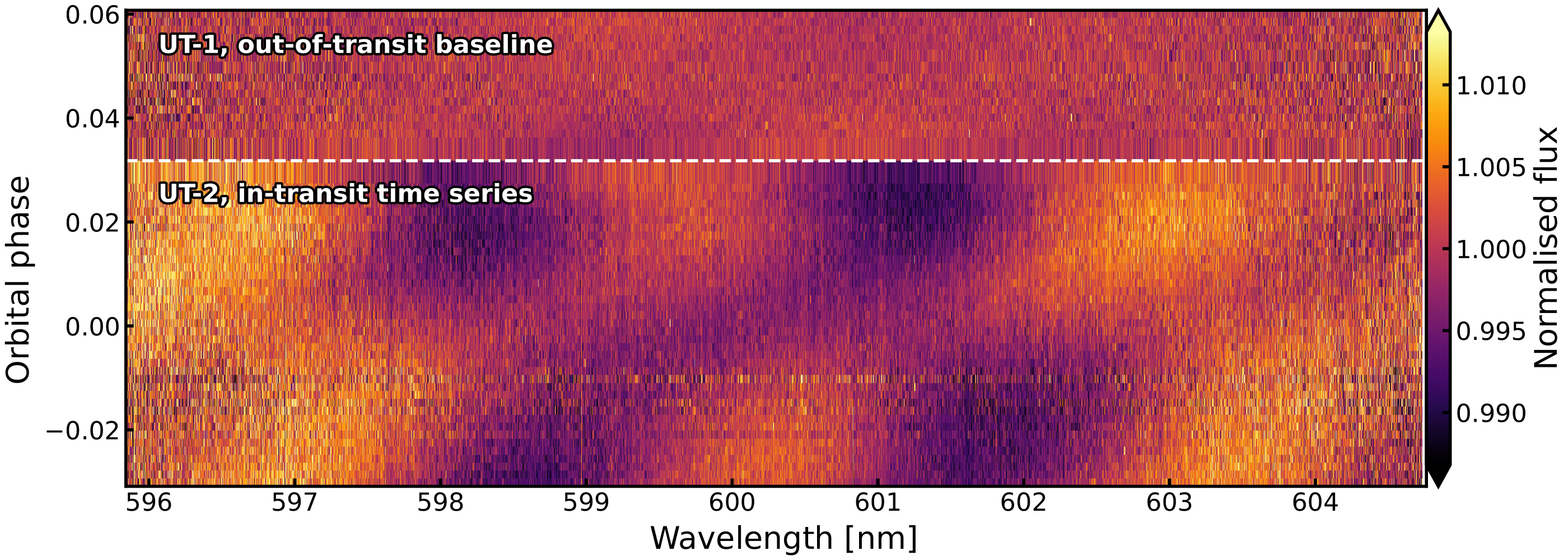}
     \caption{Normalised order showing the interference pattern caused by internal reflections in the Coudé train of ESPRESSO. The spectra have been corrected for telluric contamination, outliers have been masked out and colour correction has been applied. The order is then divided by the out-of-transit baseline. It shows that UT-1 (out of transit) is likely less affected than UT-2.}
     \label{fig:ugly_pattern}
 \end{figure*}

\begin{figure}[ht!]
    \centering
    \includegraphics[width=0.5\linewidth]{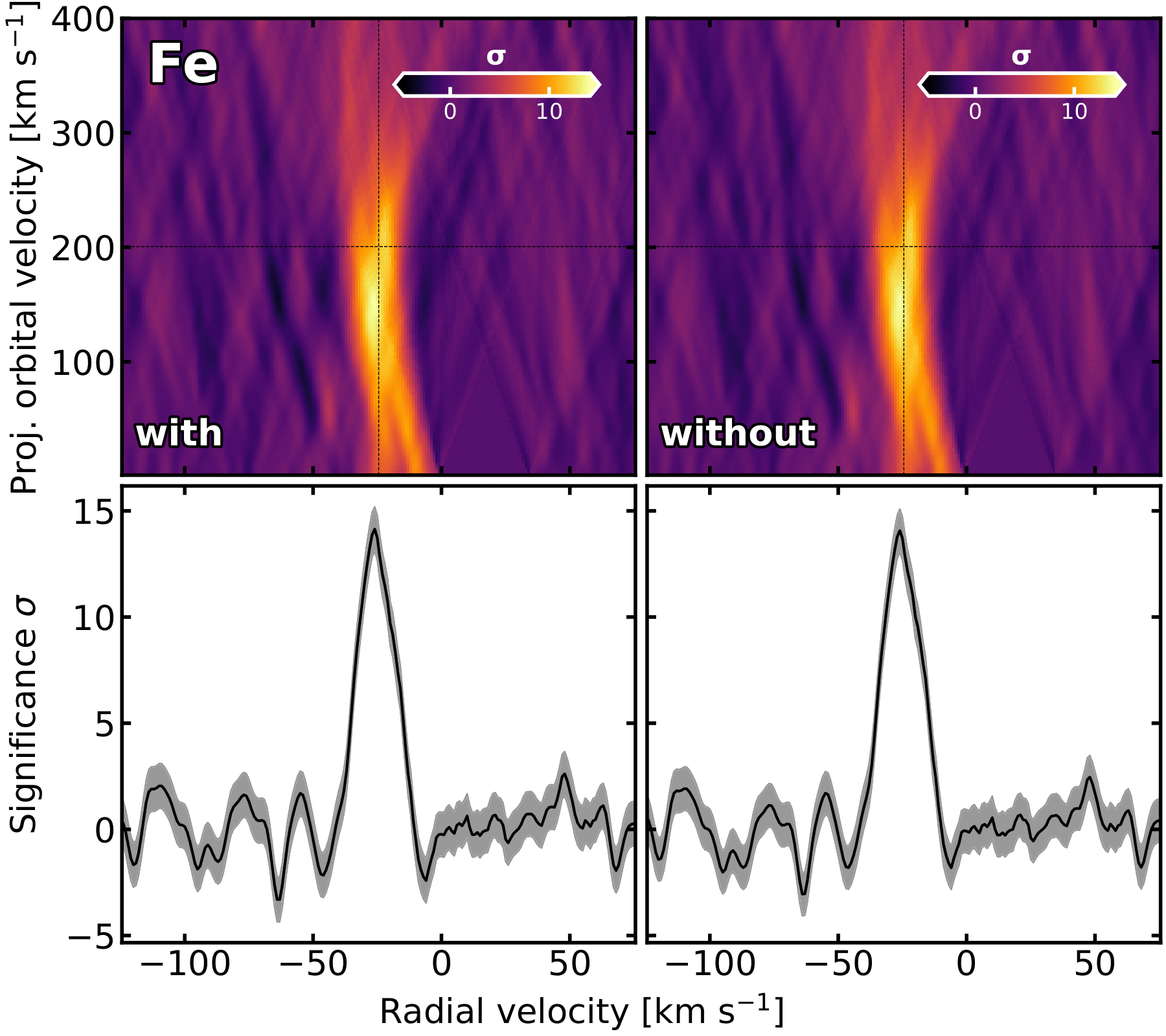}
    \caption{Comparison between cross-correlation results for the \ch{Fe} detection in the ESPRESSO transit for the sinusoidal correction in Eq. \ref{eq:sinusoid}. \textit{Left:} Cross-correlation result for \ch{Fe} applying the sinusoidal correction. \textit{Right:} Cross-correlation result for \ch{Fe} without applying the sinusoidal correction. The peak significance of the detection including the sinusoidal fit is marginally larger, but insignificantly so.}
    \label{fig:cc_comp_fe}
\end{figure}

\newpage
\section{Templates}

\begin{figure*}[ht!]
    \centering
    \includegraphics[width=\textwidth]{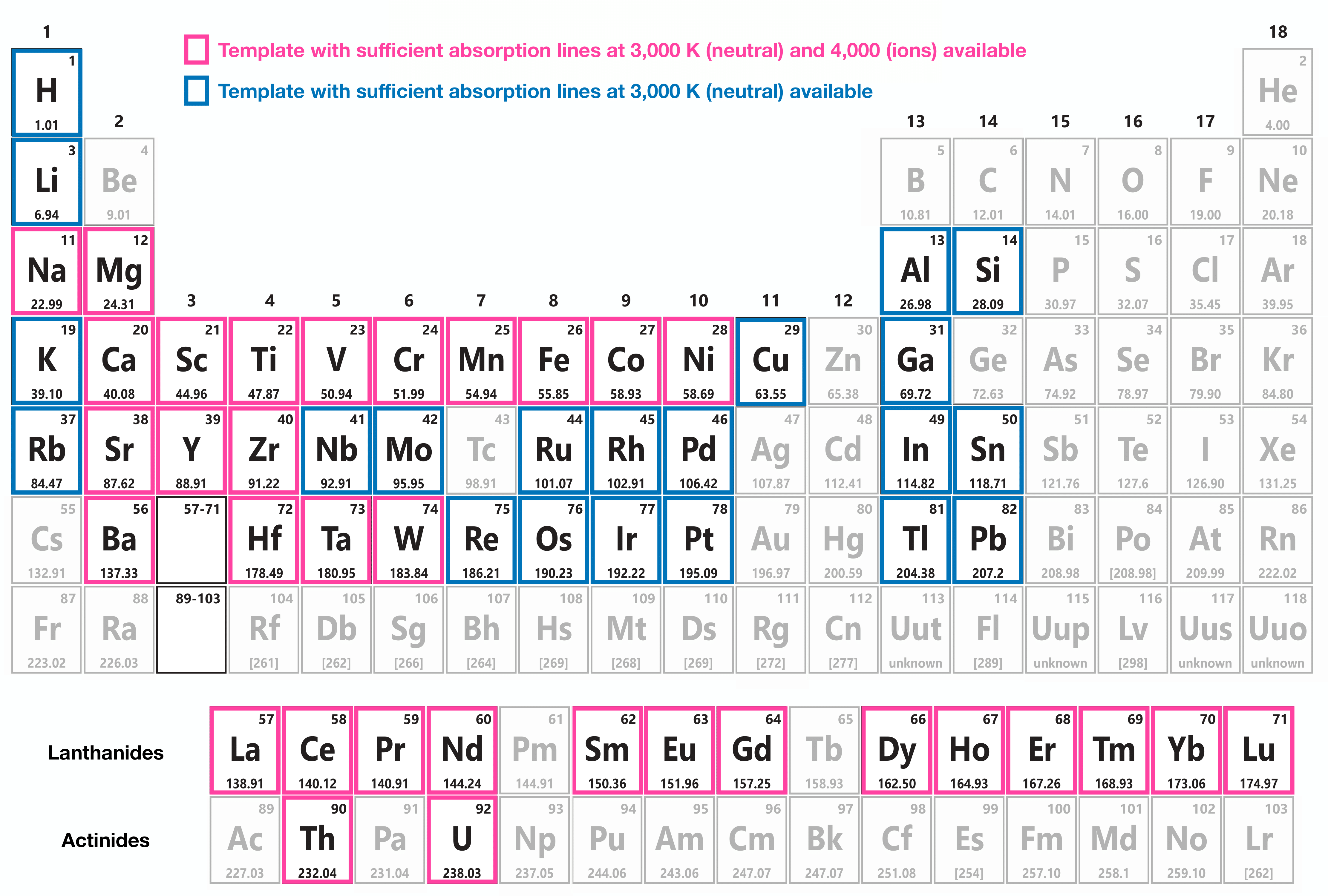}
    \caption{Periodic table of the elements indicating the considered cross-correlation templates in this work. Pink frames indicate the availability of sufficient absorption lines at \SI{3000}{\kelvin} (neutral atoms) and \SI{4000}{\kelvin} (ions) in the wavelength regions of HARPS \& HARPS-N, ESPRESSO, and MAROON-X. Blue frames indicate the availability of sufficient absorption lines at \SI{3000}{\kelvin} (neutral atoms), but not for the ionised atom at higher temperatures. Templates for \ch{TiO}, \ch{VO}, and \ch{AlO} are also available at \SI{2500}{\kelvin}. The cross-correlation analysis is performed for all of the available templates as indicated in this table and for the oxides mentioned above (90 species in total).}
    \label{fig:periodic_system}
\end{figure*}

\newpage

\section{Trace-fit posteriors and best-fit models}

\begin{figure*}[ht!]
    \centering
    \includegraphics[width=\textwidth]{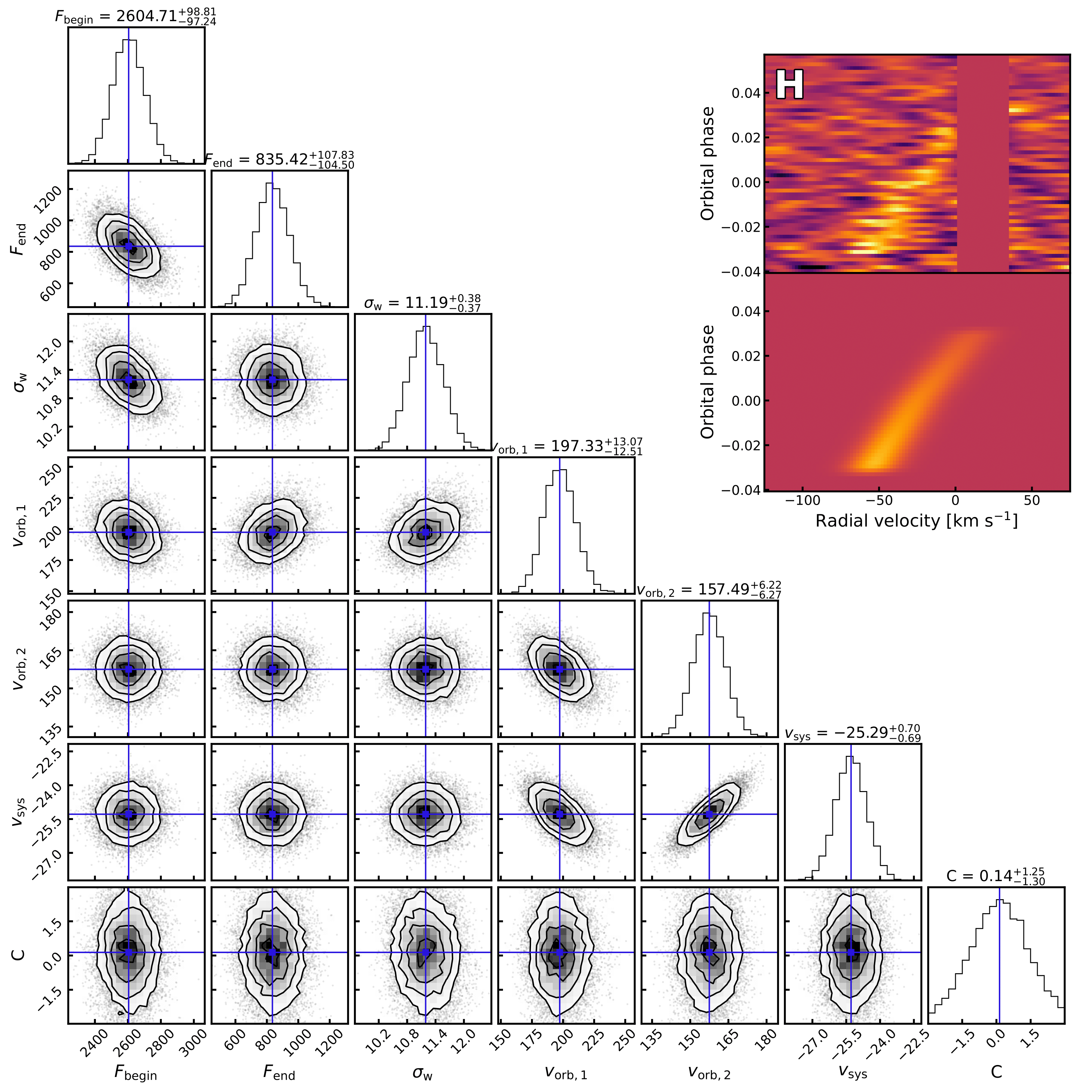}
    \caption{Posterior distributions of the model parameters of Eq.\,\ref{eq:ccf_gauss} for \ch{H}. The top-right panel shows the two-dimensional cross-correlation function of all eight transit times series combined binned to a common phase grid (top) and the model with parameters equal to the median of their posterior distributions (bottom). }
    \label{fig:trace_H}
\end{figure*}
\newpage
\begin{figure*}[ht!]
    \centering
    \includegraphics[width=\textwidth]{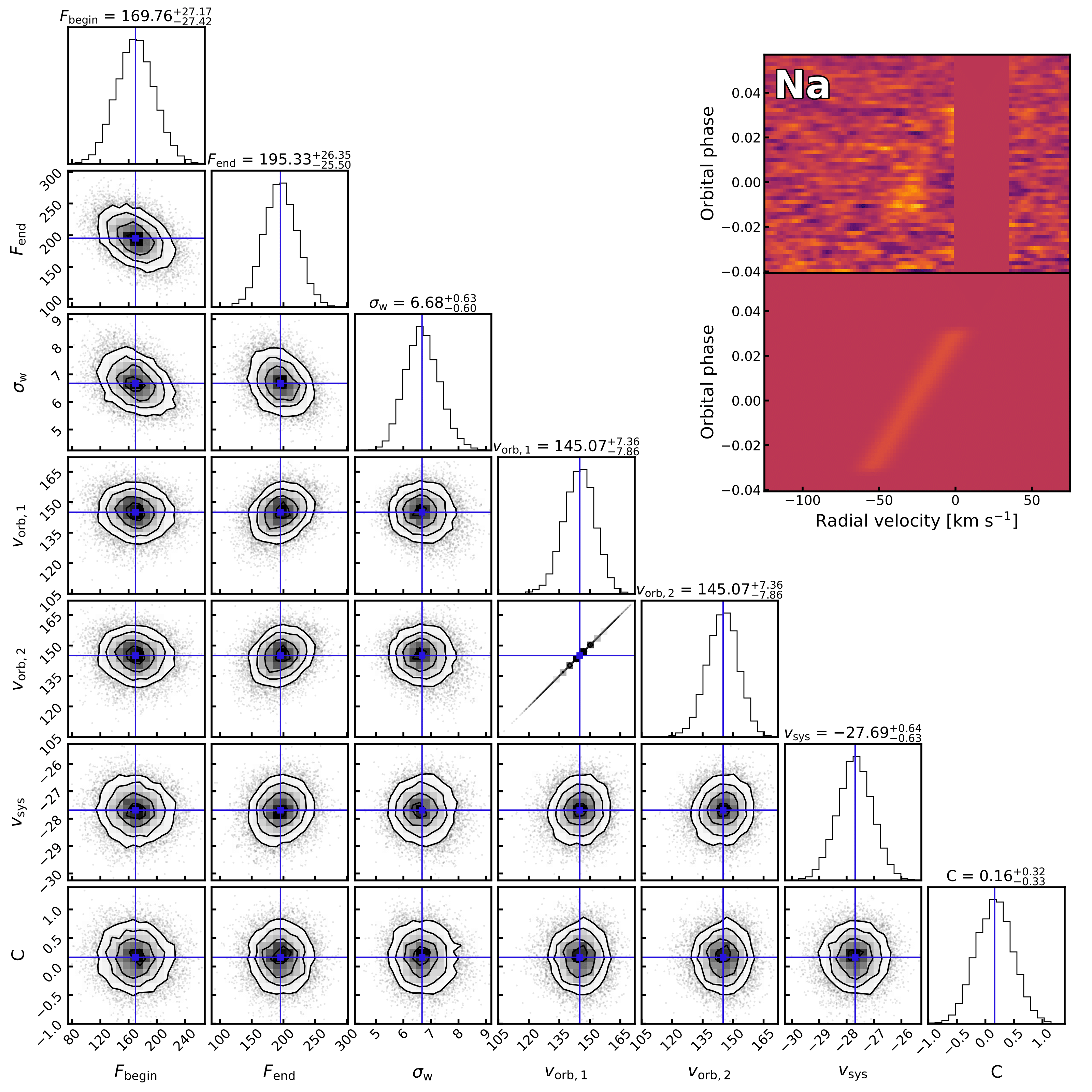}
    \caption{Same as Fig.\,\ref{fig:trace_H}, but for \ch{Na}. Note that for \ch{Na}, the projected orbital velocity was forced to be equal for both regions of the transit as the posterior distribution reached the bounds of feasible priors. To avoid correlations, we opted for fitting only one orbital velocity instead of two.}
    \label{fig:trace_Na}
\end{figure*}
\newpage
\begin{figure*}[ht!]
    \centering
    \includegraphics[width=\textwidth]{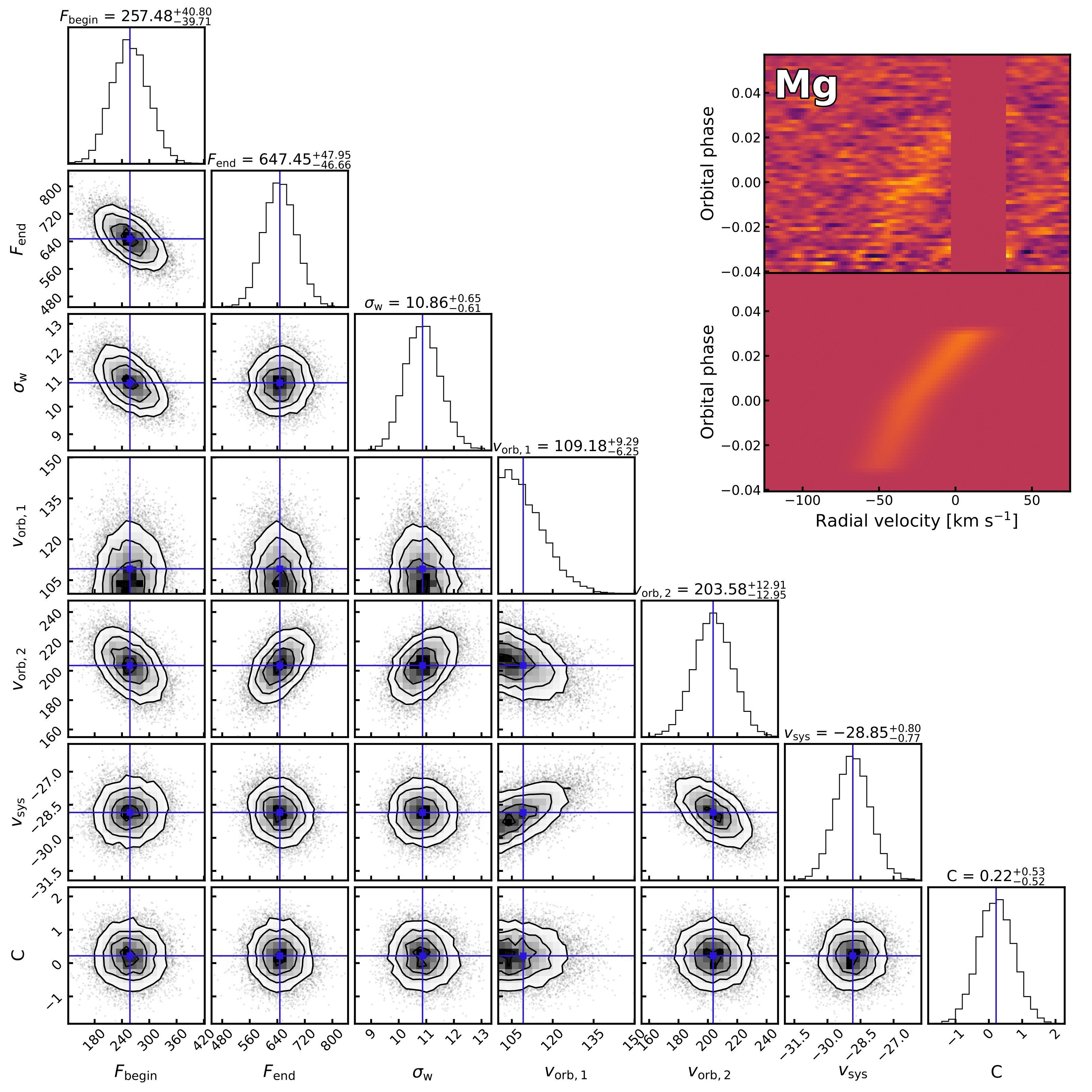}
    \caption{Same as Fig.\,\ref{fig:trace_H}, but for \ch{Mg}. Note that we allow two orbital velocities to be fit despite the weak signal in the first half of the transit.}
    \label{fig:trace_Mg}
\end{figure*}
\newpage
\begin{figure*}[ht!]
    \centering
    \includegraphics[width=\textwidth]{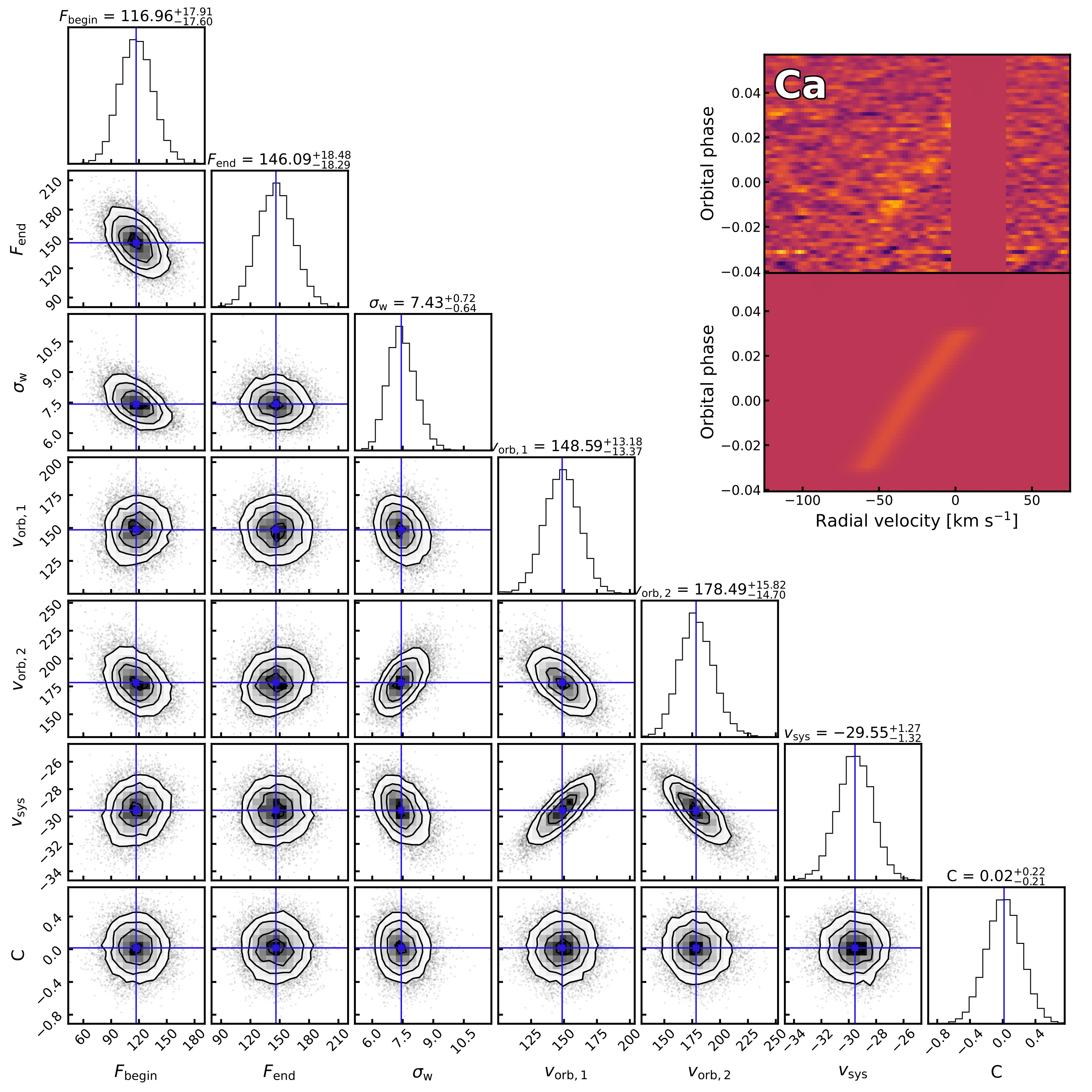}
    \caption{Same as Fig.\,\ref{fig:trace_H}, but for \ch{Ca}.}
    \label{fig:trace_Ca}
\end{figure*}
\newpage
\begin{figure*}[ht!]
    \centering
    \includegraphics[width=\textwidth]{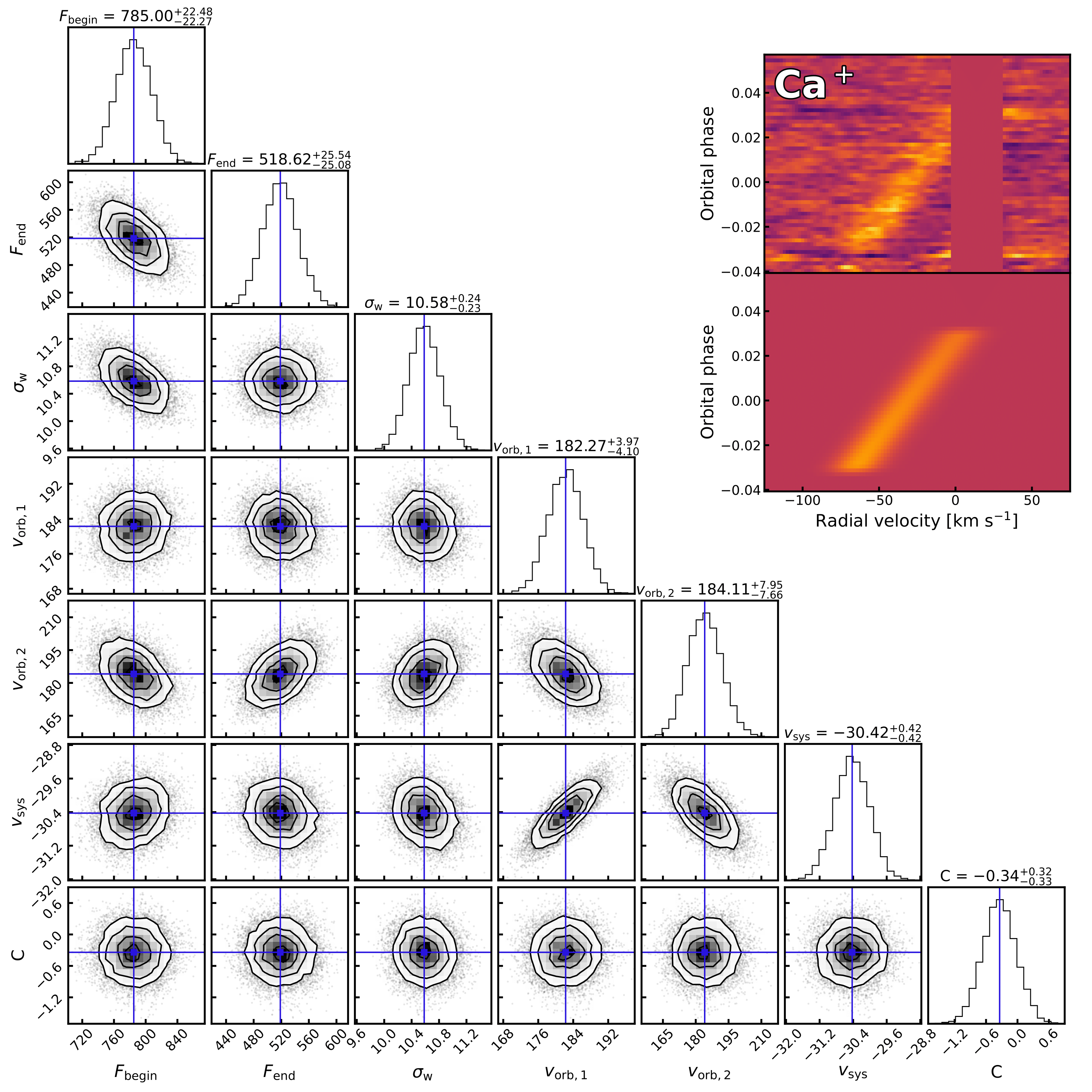}
    \caption{Same as Fig.\,\ref{fig:trace_H}, but for \ch{Ca+}.}
    \label{fig:trace_Cap}
\end{figure*}

\newpage
\begin{figure*}[ht!]
    \centering
    \includegraphics[width=\textwidth]{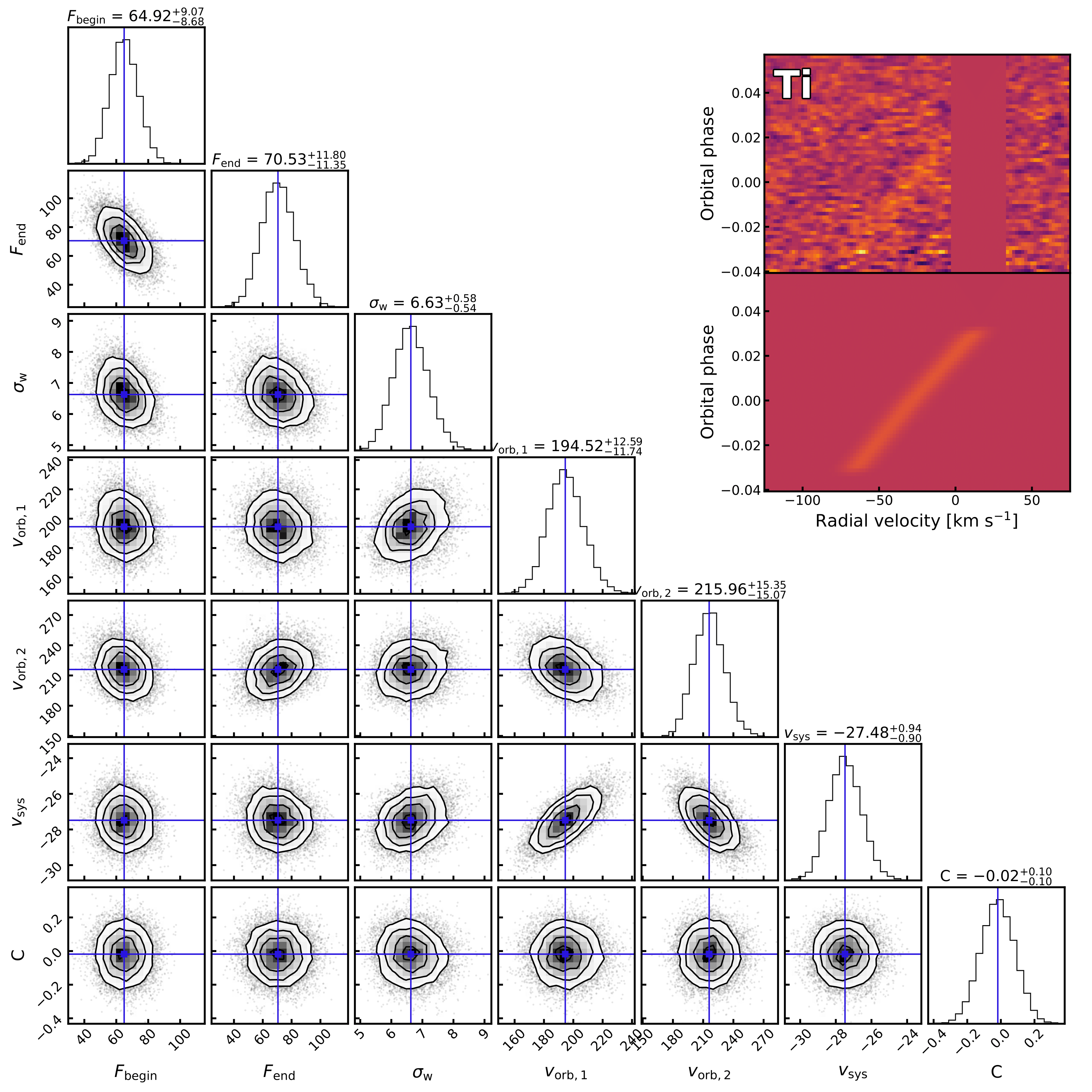}
    \caption{Same as Fig.\,\ref{fig:trace_H}, but for \ch{Ti}.}
    \label{fig:trace_Ti}
\end{figure*}
\newpage
\begin{figure*}[ht!]
    \centering
    \includegraphics[width=\textwidth]{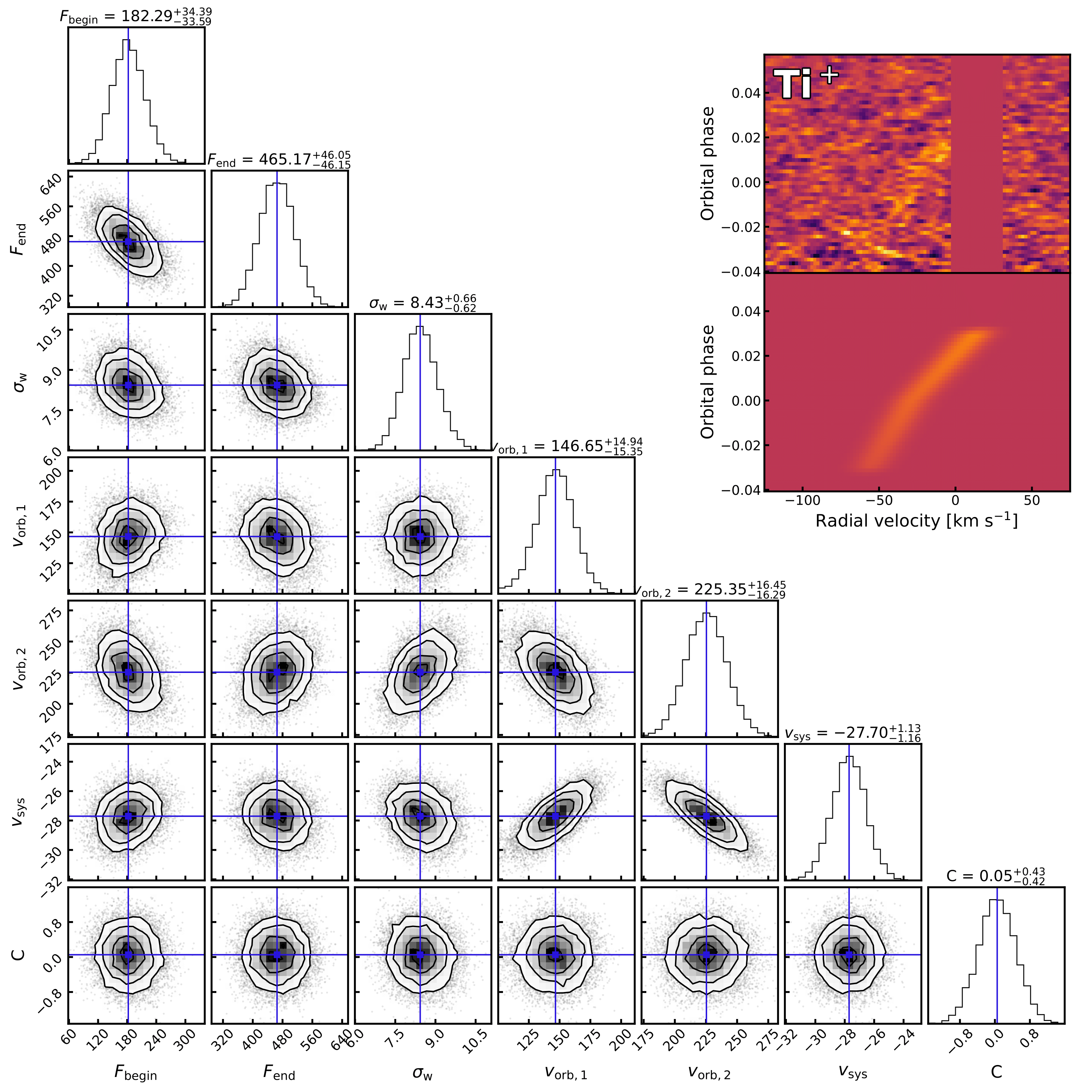}
    \caption{Same as Fig.\,\ref{fig:trace_H}, but for \ch{Ti+}.}
    \label{fig:trace_Tip}
\end{figure*}
\newpage
\begin{figure*}[ht!]
    \centering
    \includegraphics[width=\textwidth]{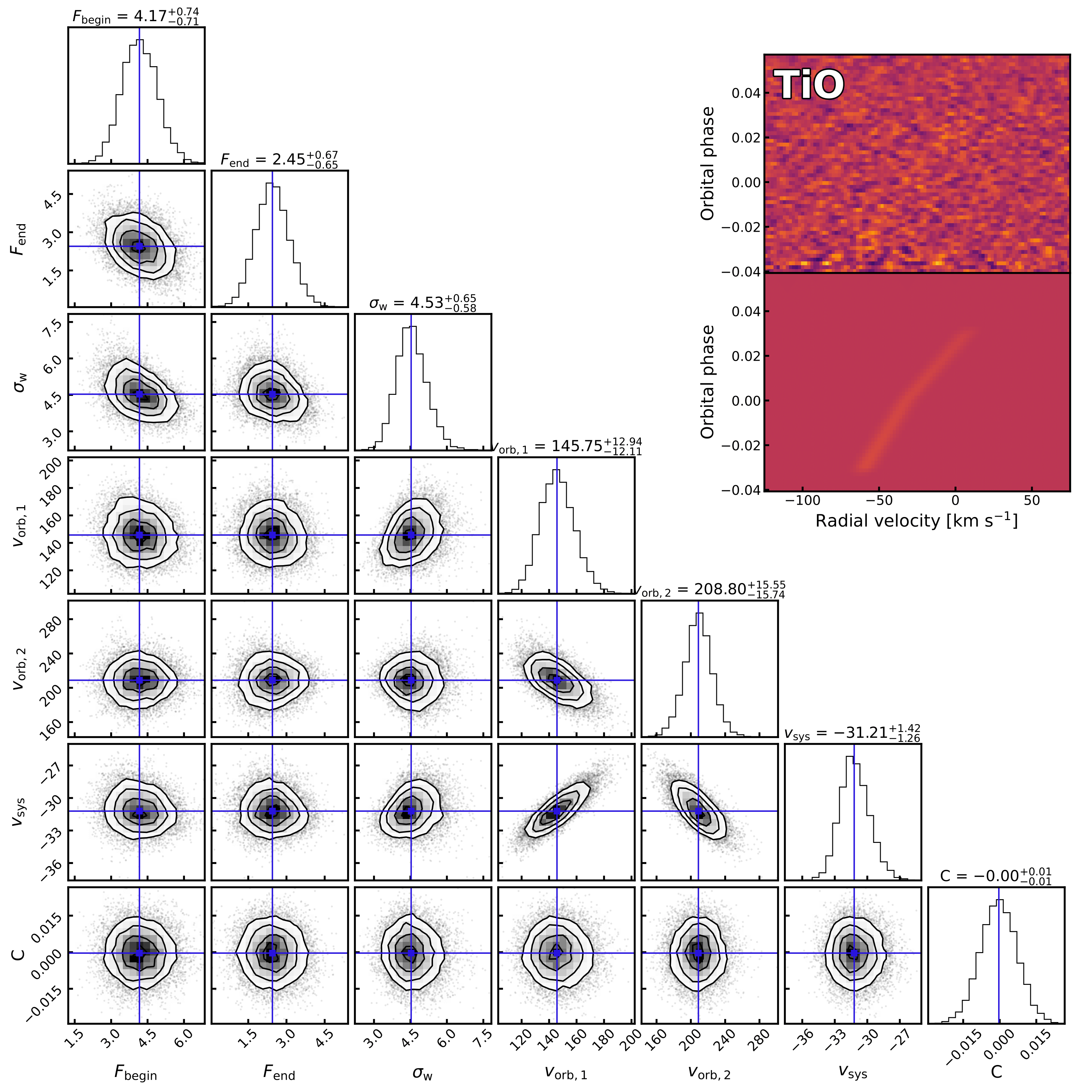}
    \caption{Same as Fig.\,\ref{fig:trace_H}, but for \ch{TiO}.}
    \label{fig:trace_TiO}
\end{figure*}
\newpage
\begin{figure*}[ht!]
    \centering
    \includegraphics[width=\textwidth]{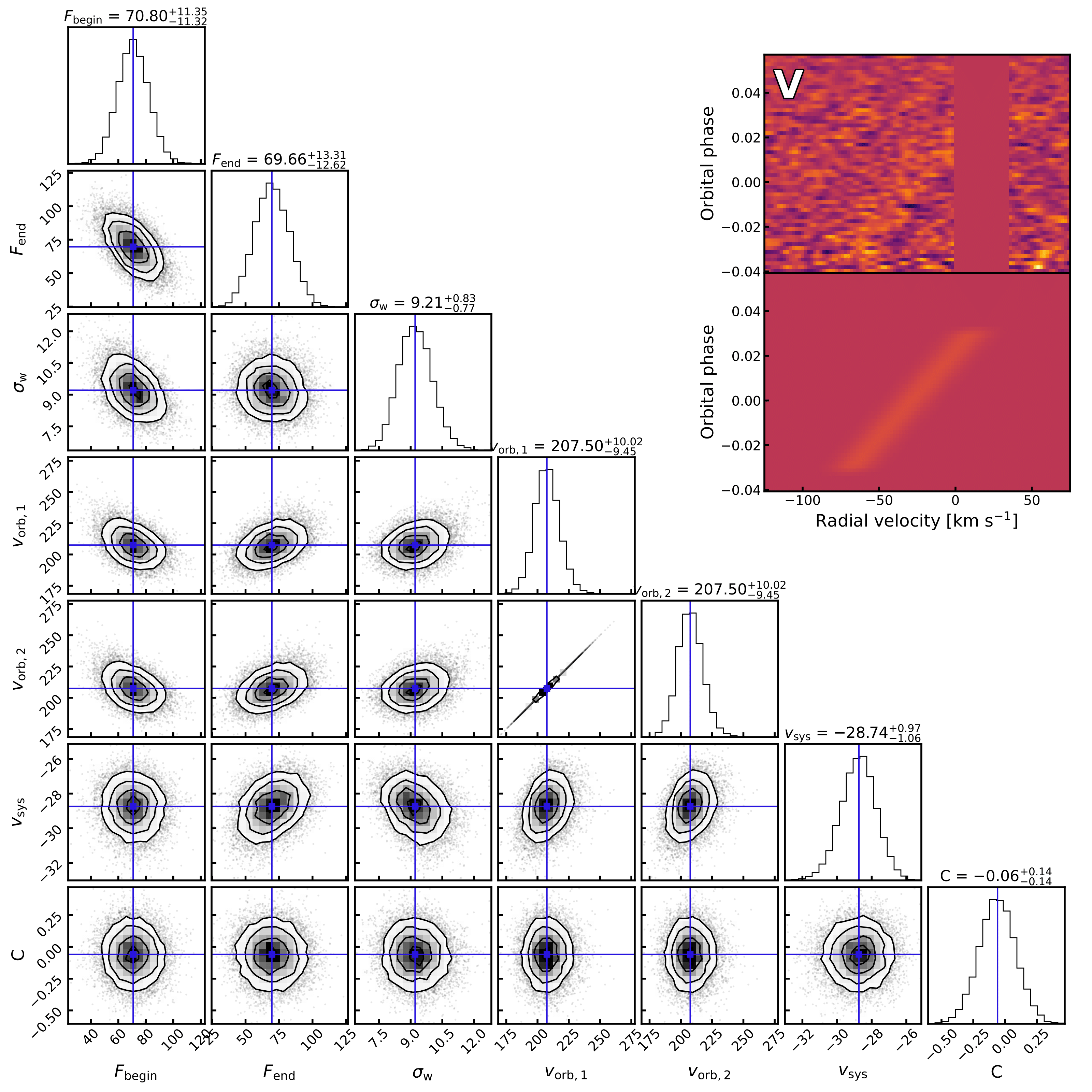}
    \caption{Same as Fig.\,\ref{fig:trace_H}, but for \ch{V}. Note that we allow two orbital velocities to be fit despite the weak signal in the first half of the transit.}
    \label{fig:trace_V}
\end{figure*}
\newpage
\begin{figure*}[ht!]
    \centering
    \includegraphics[width=\textwidth]{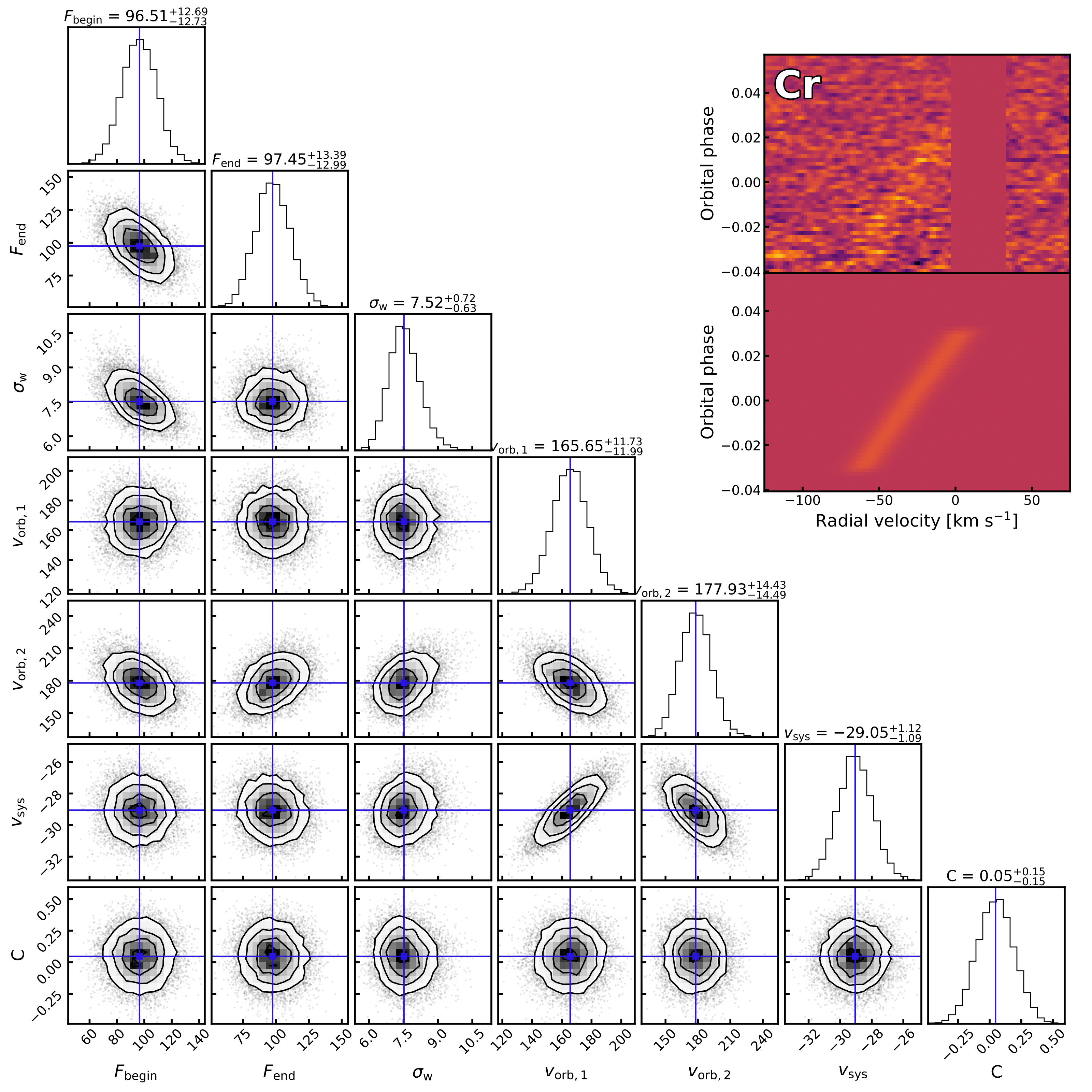}
    \caption{Same as Fig.\,\ref{fig:trace_H}, but for \ch{Cr}.}
    \label{fig:trace_Cr}
\end{figure*}

\newpage
\begin{figure*}[ht!]
    \centering
    \includegraphics[width=\textwidth]{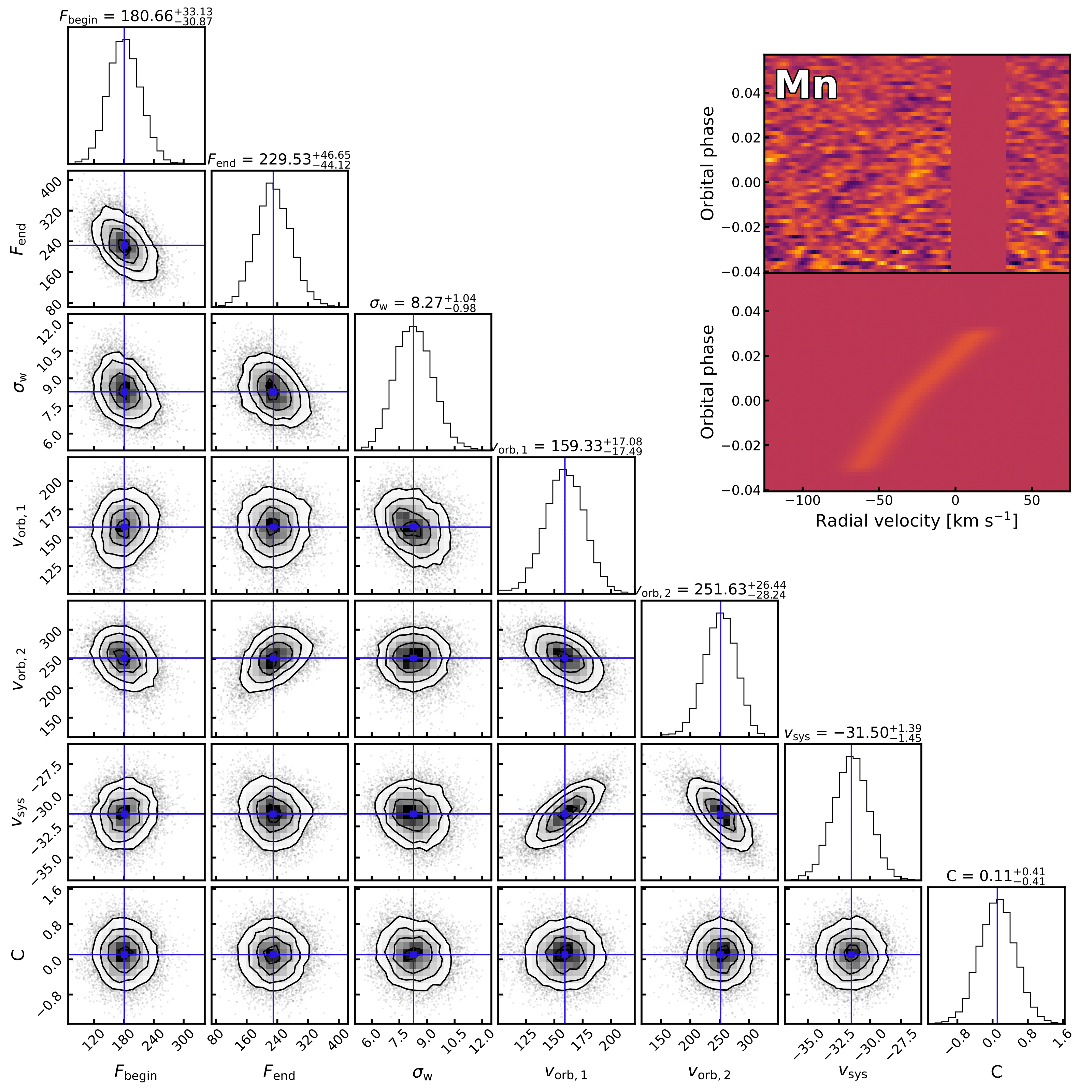}
    \caption{Same as Fig.\,\ref{fig:trace_H}, but for \ch{Mn}.}
    \label{fig:trace_Mn}
\end{figure*}
\newpage
\begin{figure*}[ht!]
    \centering
    \includegraphics[width=\textwidth]{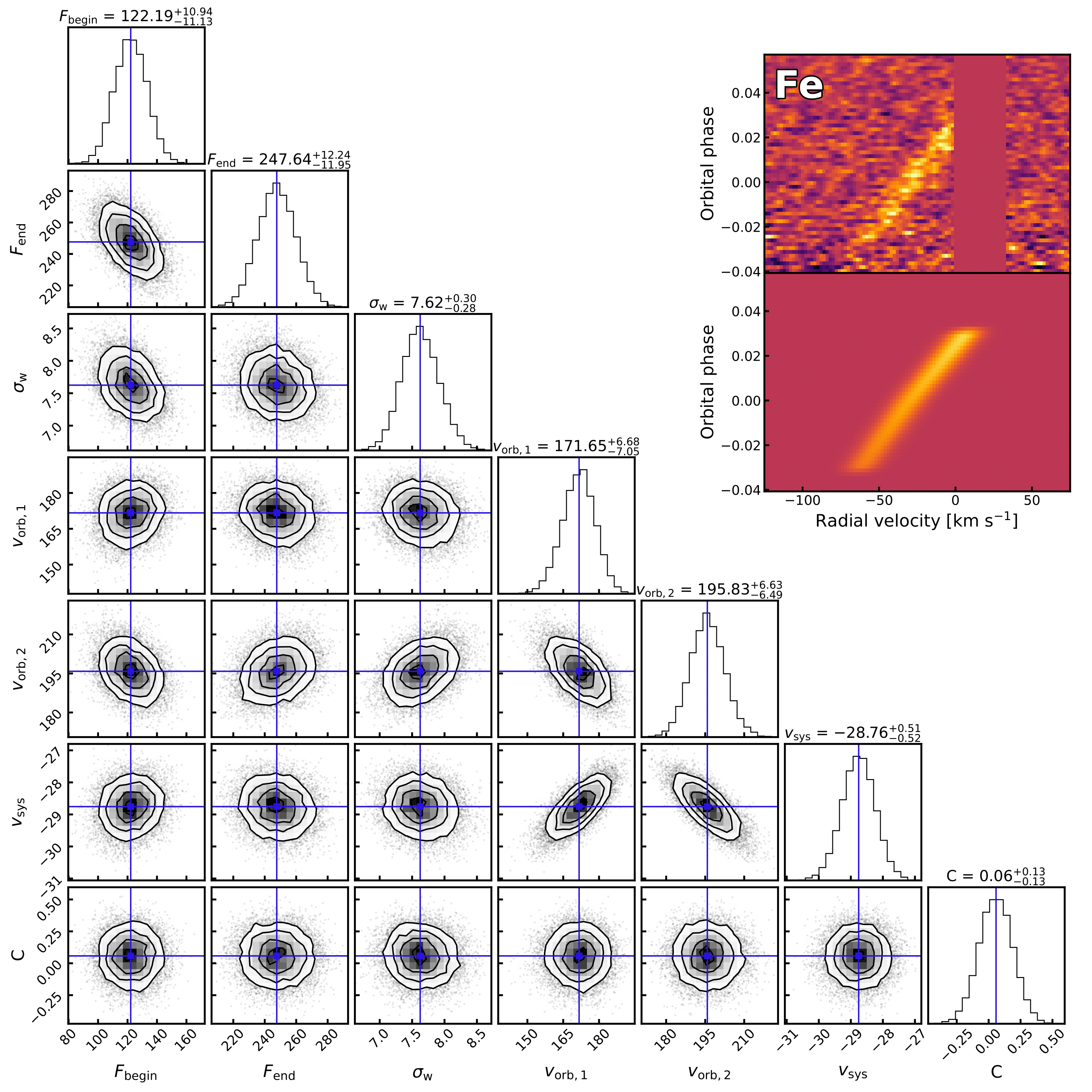}
    \caption{Same as Fig.\,\ref{fig:trace_H}, but for \ch{Fe}.}
    \label{fig:trace_Fe}
\end{figure*}
\newpage
\begin{figure*}[ht!]
    \centering
    \includegraphics[width=\textwidth]{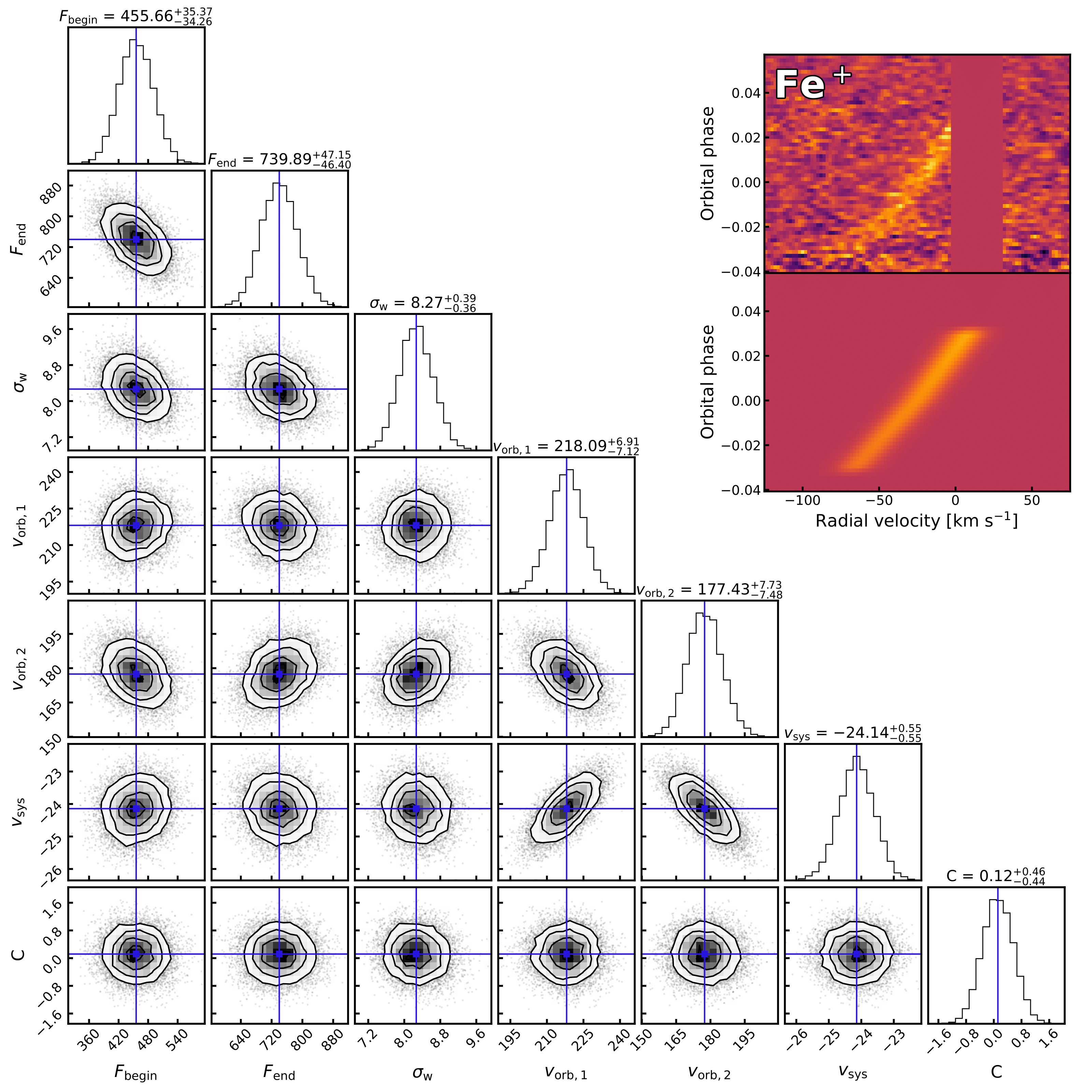}
    \caption{Same as Fig.\,\ref{fig:trace_H}, but for \ch{Fe+}.}
    \label{fig:trace_Fep}
\end{figure*}
\newpage
\begin{figure*}[ht!]
    \centering
    \includegraphics[width=\textwidth]{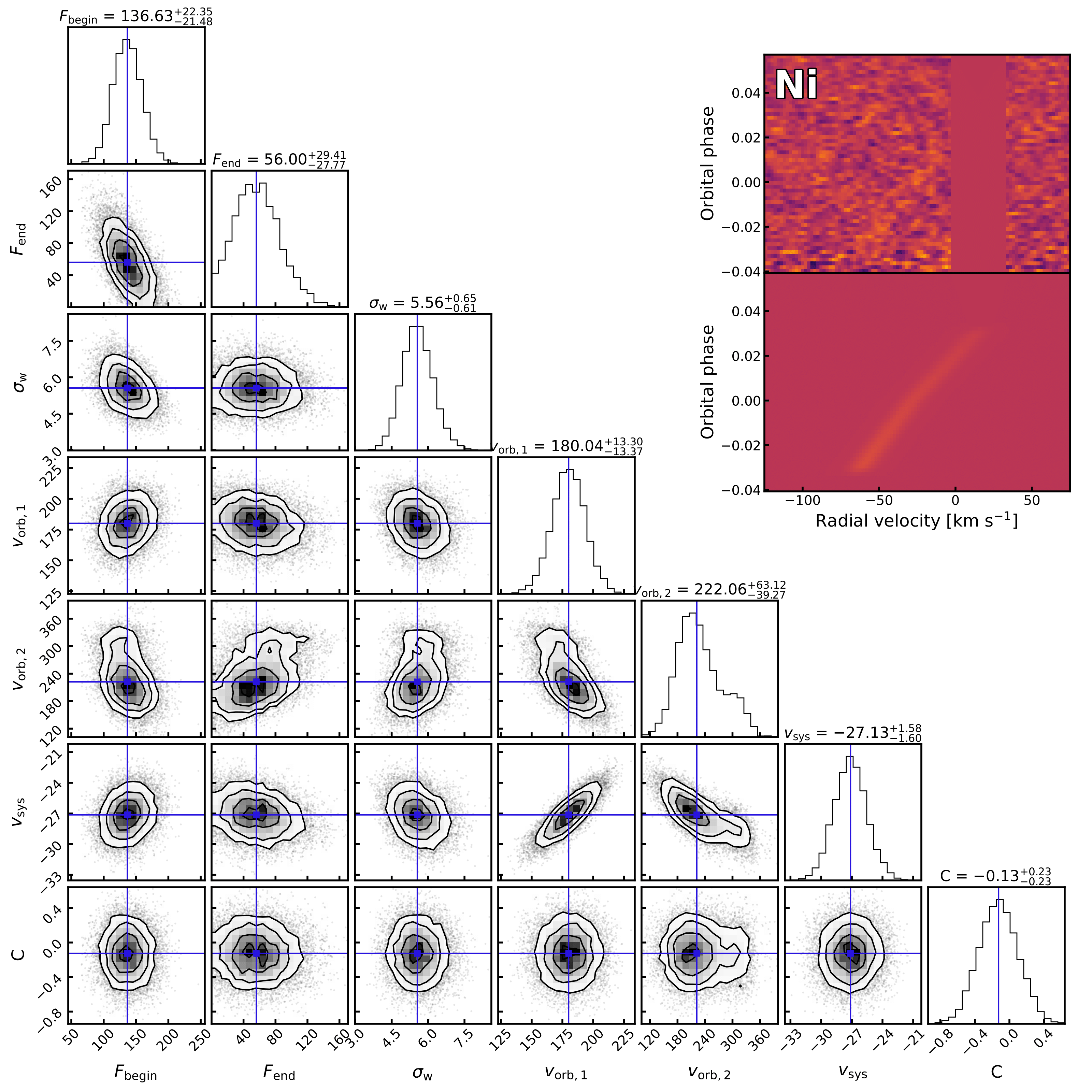}
    \caption{Same as Fig.\,\ref{fig:trace_H}, but for \ch{Ni}.}
    \label{fig:trace_Ni}
\end{figure*}
\newpage
\begin{figure*}[ht!]
    \centering
    \includegraphics[width=\textwidth]{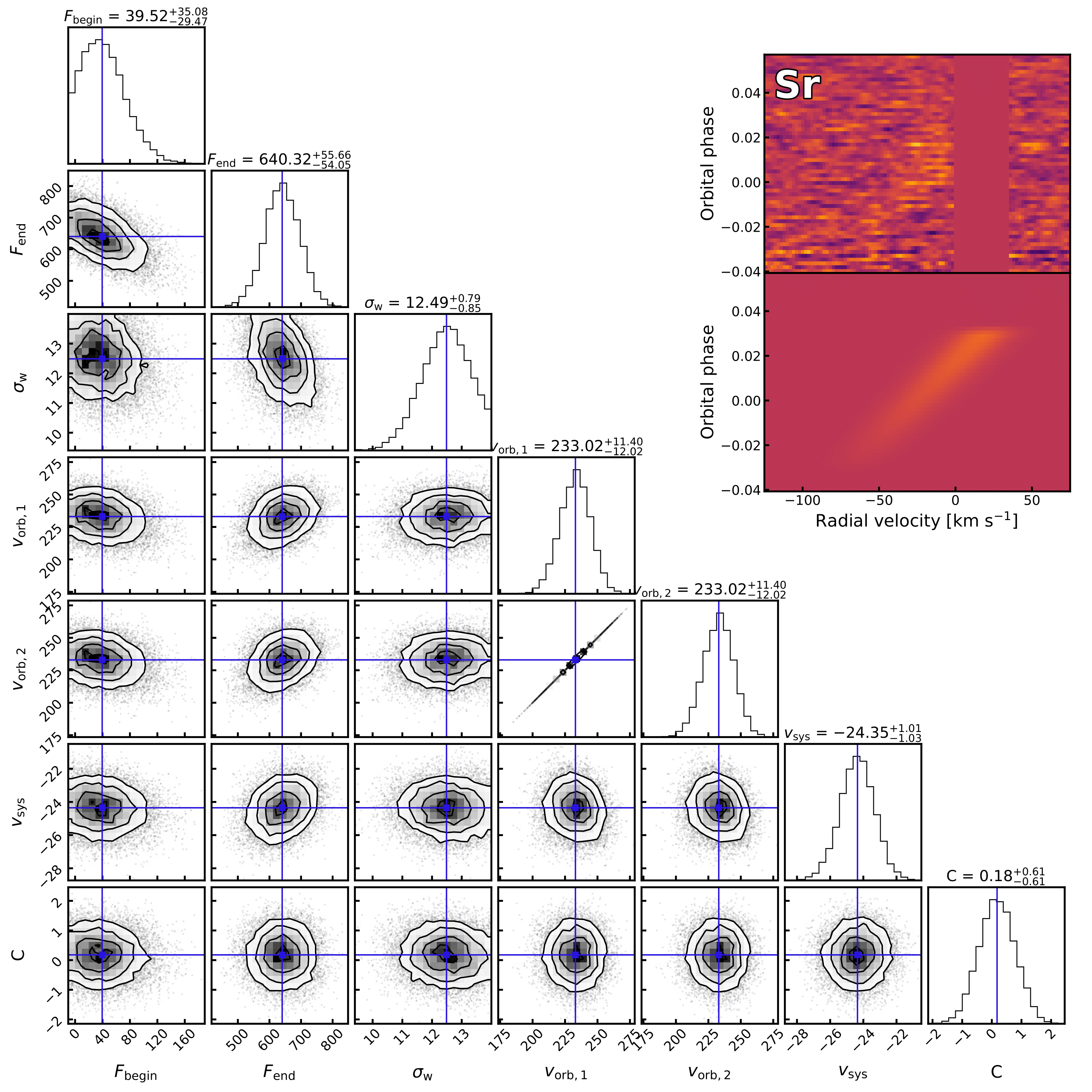}
    \caption{Same as Fig.\,\ref{fig:trace_H}, but for \ch{Sr}. Note that the orbital velocity was forced to be the same in both halves of the transit.}
    \label{fig:trace_Sr}
\end{figure*}
\newpage
\begin{figure*}[ht!]
    \centering
    \includegraphics[width=\textwidth]{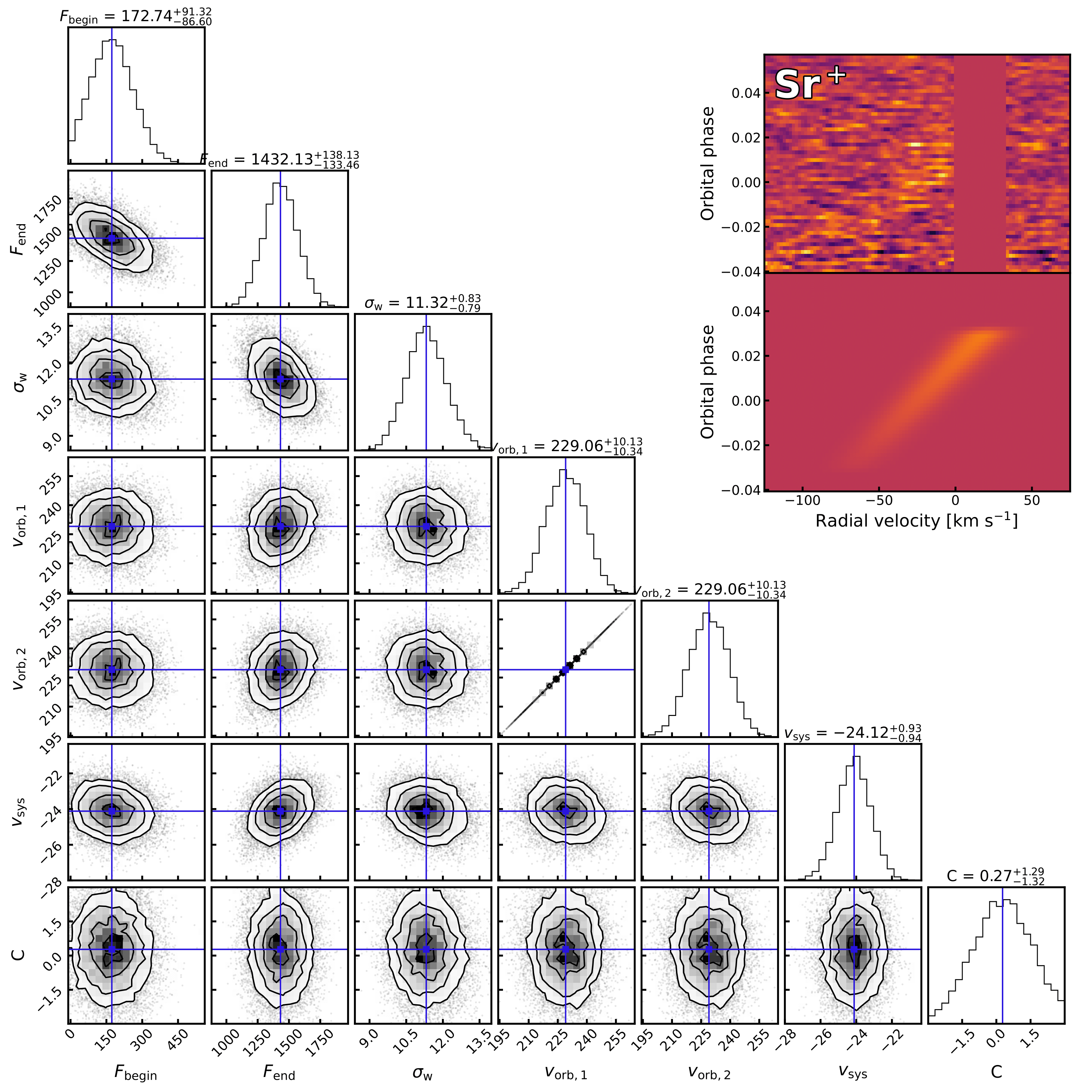}
    \caption{Same as Fig.\,\ref{fig:trace_H}, but for \ch{Sr+}. Note that the orbital velocity was forced to be the same in both halves of the transit.}
    \label{fig:trace_Srp}
\end{figure*}

\newpage
\begin{figure*}[ht!]
    \centering
    \includegraphics[width=\textwidth]{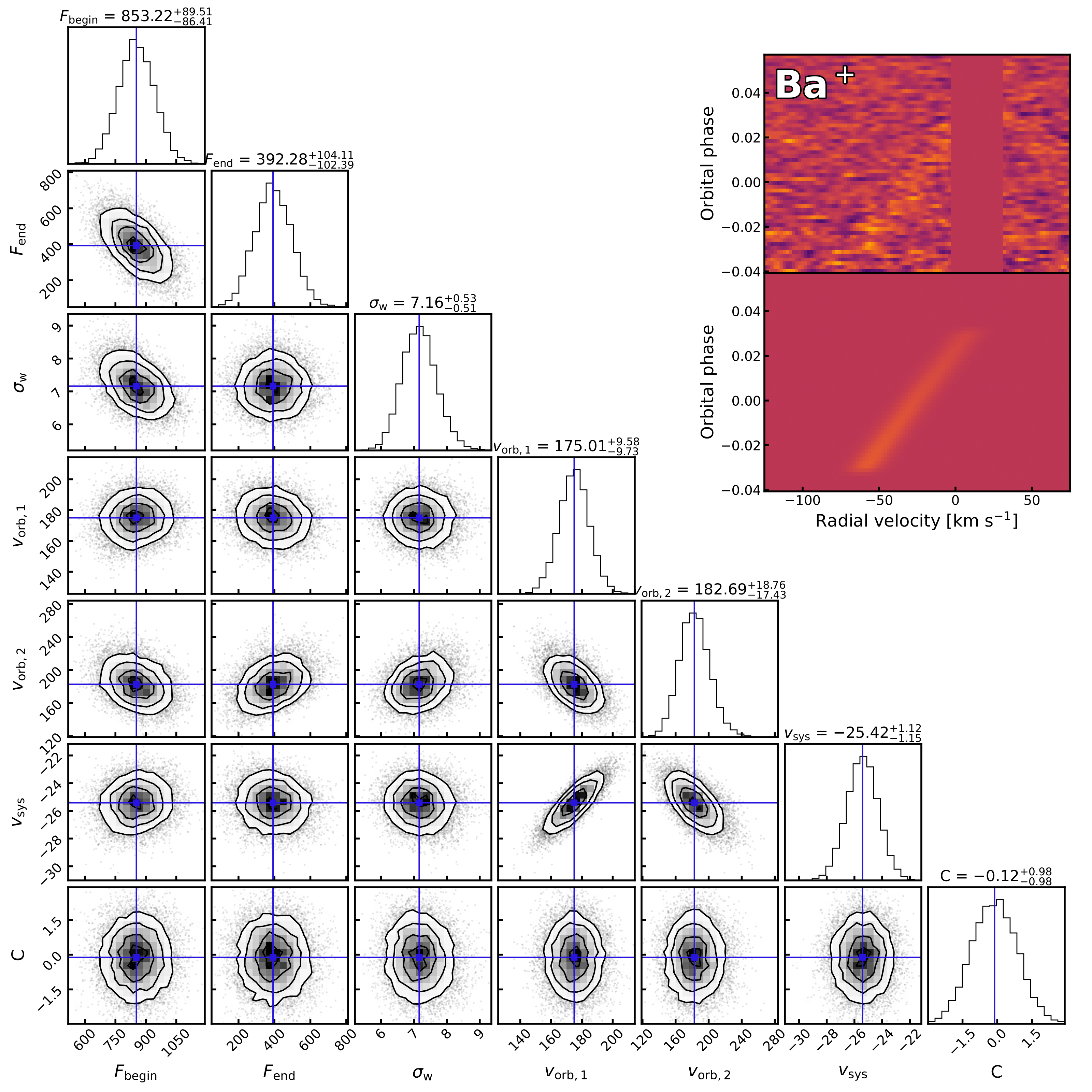}
    \caption{Same as Fig.\,\ref{fig:trace_H}, but for \ch{Ba+}.}
    \label{fig:Ba_p_trace}
\end{figure*}

\newpage
\begin{figure*}[ht!]
    \centering
    \includegraphics[width=\textwidth]{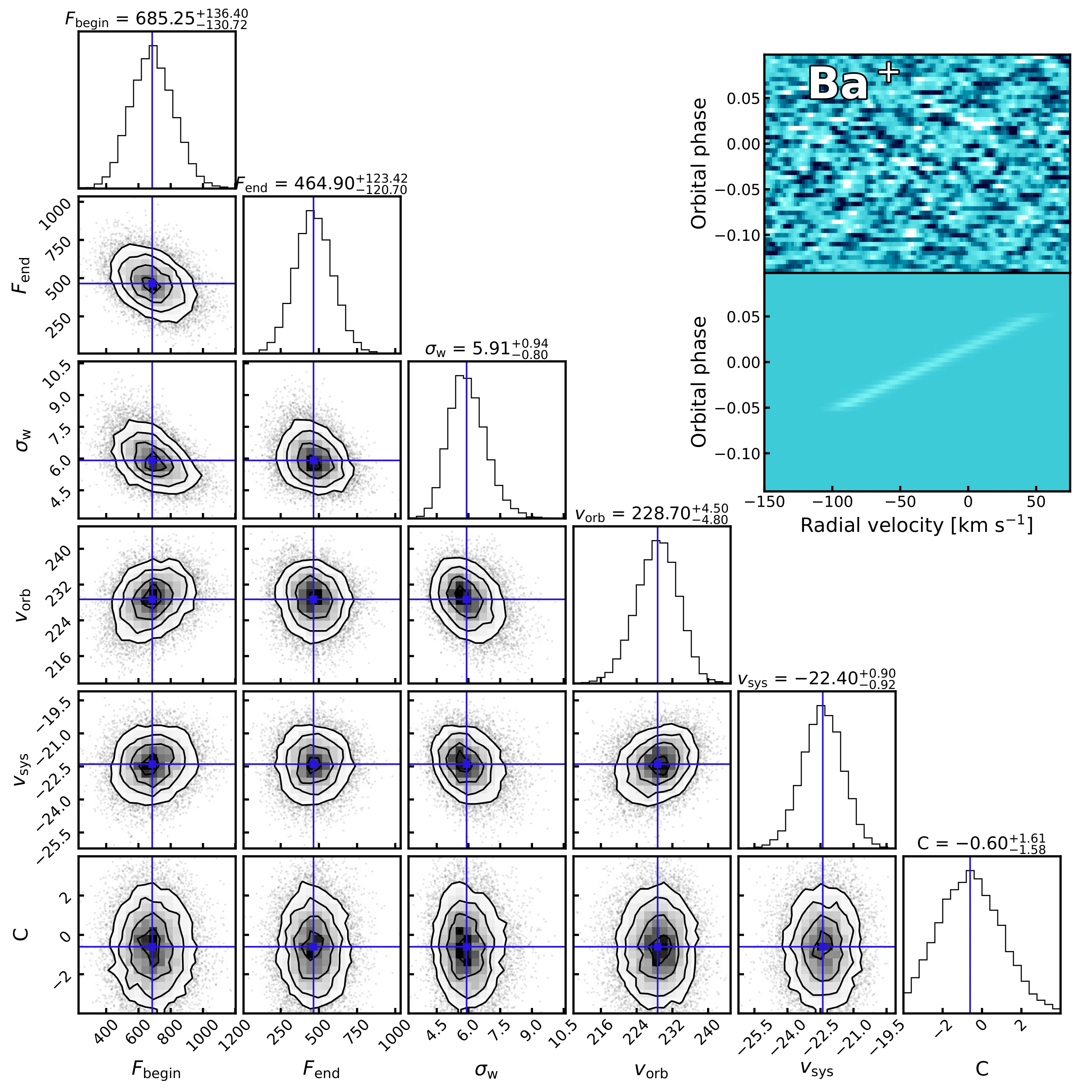}
    \caption{Same as Fig.\,\ref{fig:trace_H}, but for \ch{Ba+} in the atmosphere of \nic. The signal strength at the end of the transit is weaker than at the start, which agrees with the prediction that \ch{Ba+} should ionise to \ch{Ba^{2+}} on the hot dayside. Note that we only fit for one orbital velocity.}
    \label{fig:trace_Ba_nic}
\end{figure*}

\newpage
\section{Non-detections}
\begin{figure*}[ht!]
    \centering
    \includegraphics[width=\textwidth]{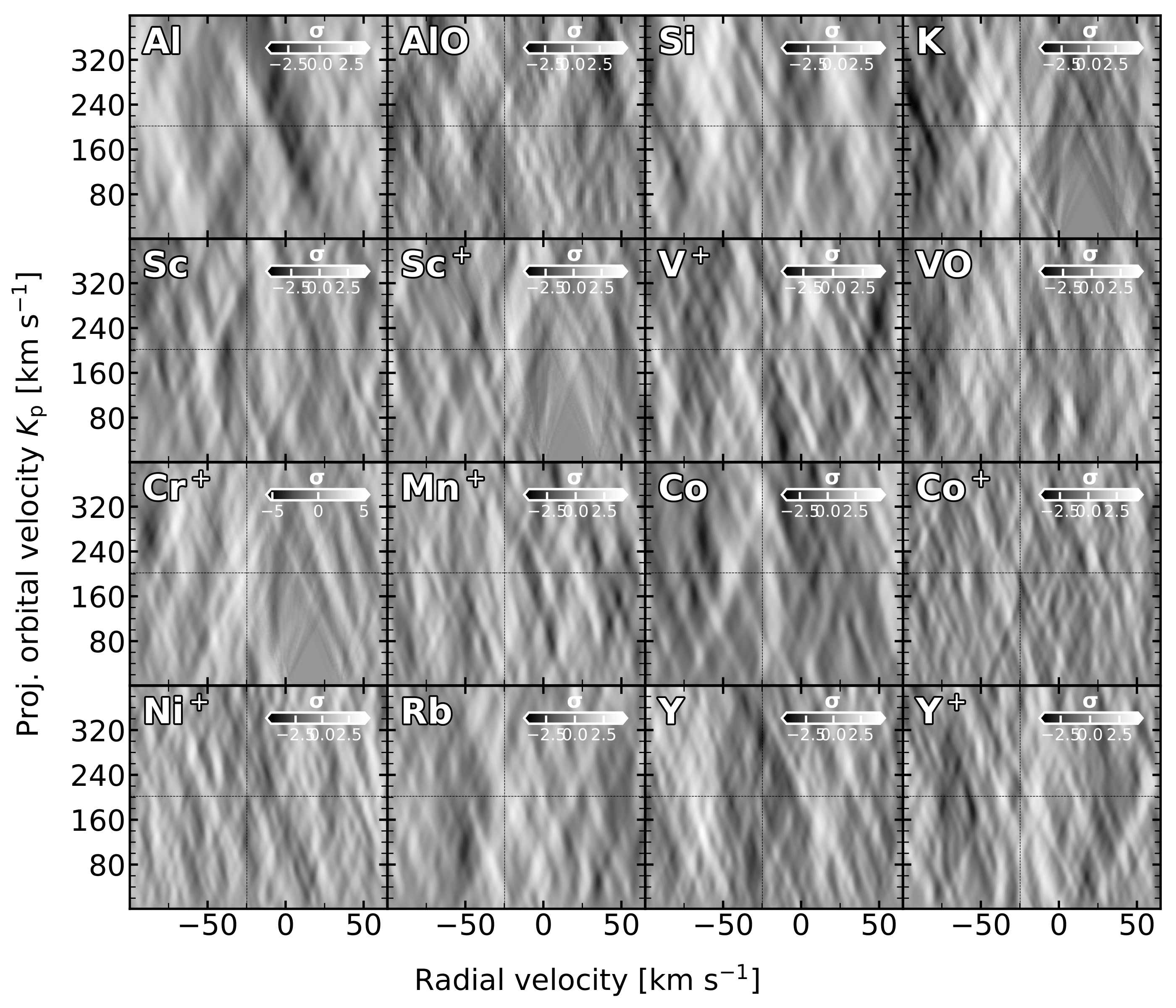}
    \caption{Same as Fig.\,\ref{fig:det}, but for non-detections.}
    \label{fig:non_det_1}
\end{figure*}
\begin{figure*}[ht!]
    \centering
    \includegraphics[width=\textwidth]{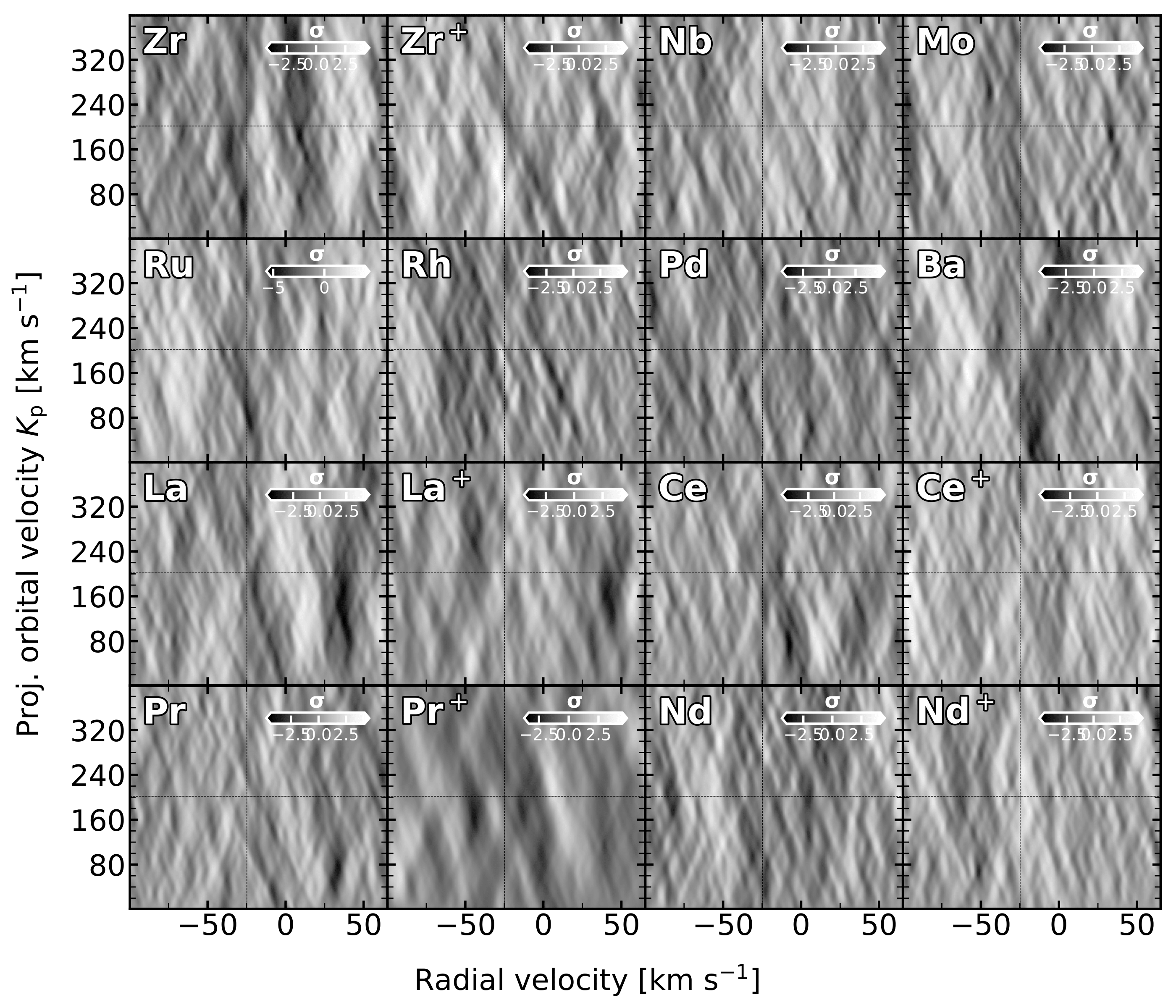}
    \caption{Same as Fig.\,\ref{fig:det}, but for non-detections.}
    \label{fig:non_det_2}
\end{figure*}

\begin{figure*}[ht!]
    \centering
    \includegraphics[width=\textwidth]{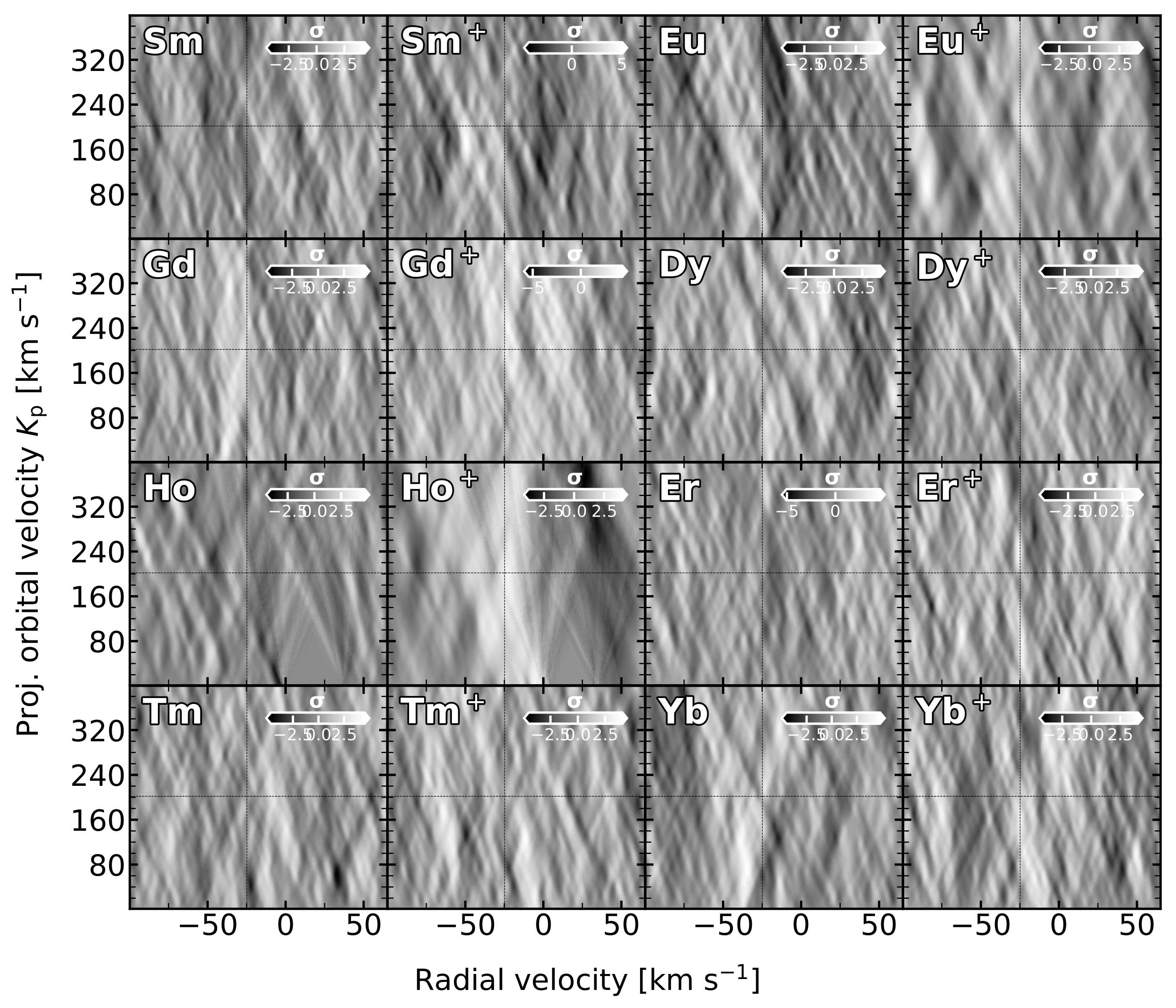}
    \caption{Same as Fig.\,\ref{fig:det}, but for non-detections.}
    \label{fig:non_det_3}
\end{figure*}

\begin{figure*}[ht!]
    \centering
    \includegraphics[width=\textwidth]{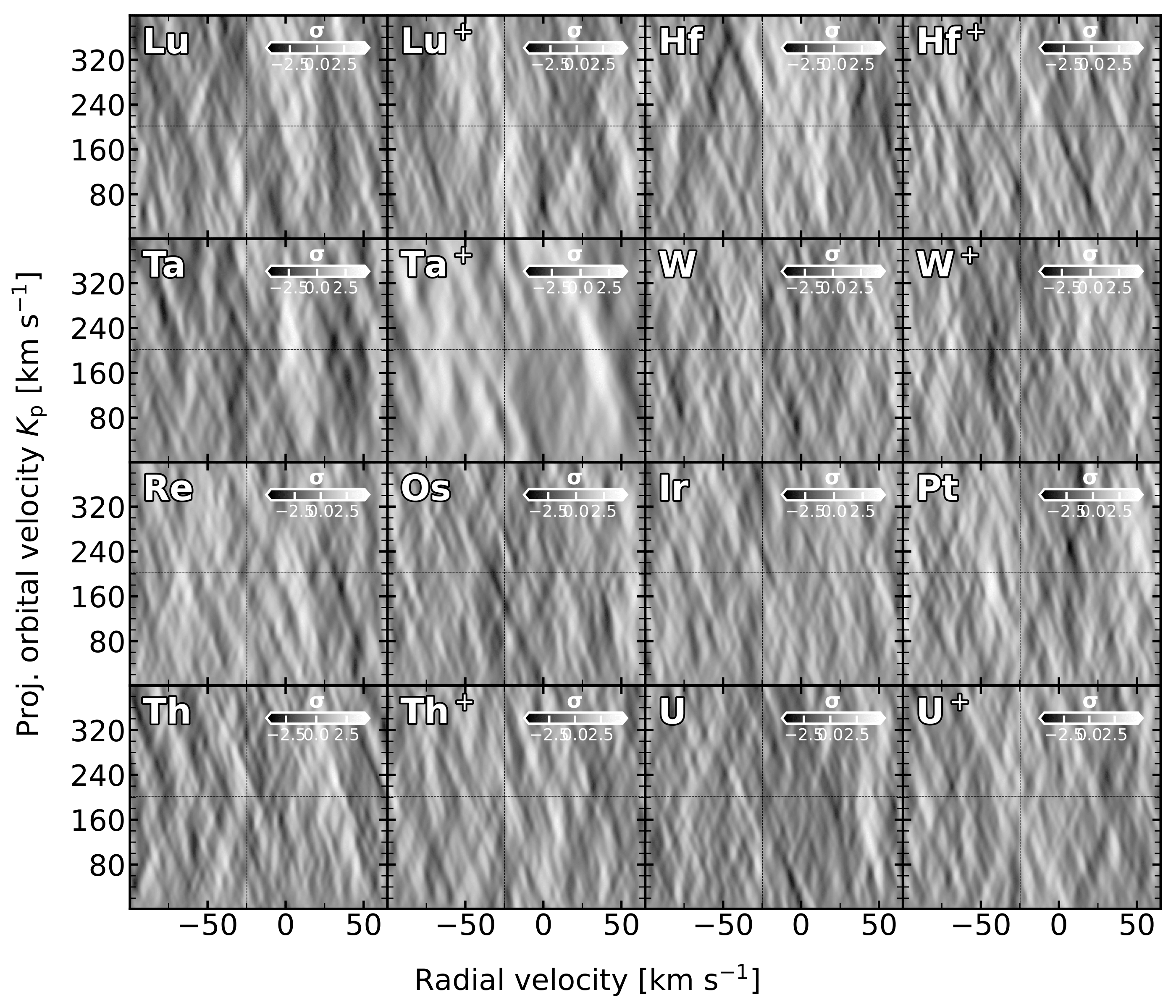}
    \caption{Same as Fig.\,\ref{fig:det}, but for non-detections.}
    \label{fig:non_det_4}
\end{figure*}

\end{document}